\documentclass[longauth]{aa}

\newcommand{\Msun}{\ensuremath{{\rm M}_\odot}}                  
\newcommand{\Rsun}{\ensuremath{{\rm R}_\odot}}                  
\newcommand{\Teff}{\ensuremath{{\rm T}_{\rm eff}}}                      
\newcommand{\logg}{\ensuremath{\log g}}                           
\newcommand{\as}{\ensuremath{^{\prime\prime}}}                    
\newcommand{\degrees}{\ensuremath{^\circ}}                        

\newcommand{\reff}[1]{{#1}}

\usepackage{graphicx}
\usepackage{txfonts}
\usepackage{natbib}

\usepackage{color}
\usepackage{epstopdf}
\usepackage{longtable}

\bibpunct{(}{)}{;}{a}{}{,} 

\begin{document}

\title{High-resolution Imaging of Transiting Extrasolar Planetary systems (HITEP).}
	\subtitle{I. Lucky imaging observations of 101 systems in the southern hemisphere\thanks{Based on data collected by the MiNDSTEp consortium using the Danish 1.54 m telescope at the ESO La Silla observatory.}\,\thanks{Tables 1, 4, and 8 are only available in electronic form at the CDS via anonymous ftp to cdsarc.u-strasbg.fr (130.79.128.5) or via http://cdsweb.u-strasbg.fr/cgi-bin/qcat?J/A+A/}}
\author{D.\ F.\ Evans \inst{\ref{keele}}
	\and J.\ Southworth \inst{\ref{keele}}
	\and P.\ F.\ L.\ Maxted \inst{\ref{keele}} 
	\and J.\ Skottfelt \inst{\ref{opencei},\ref{nbicsp}} 
	\and M.\ Hundertmark \inst{\ref{nbicsp}}
	\and U.\ G.\ J{\o}rgensen \inst{\ref{nbicsp}}
	\and M.\ Dominik \inst{\ref{sta}}
	\and K.\ A.\ Alsubai \inst{\ref{qeeri}}
	\and M.\ I.\ Andersen \inst{\ref{nbidcc}}
	\and V.\ Bozza \inst{\ref{salerno}, \ref{infn}}
	\and D.\ M.\ Bramich \inst{\ref{qeeri}}
	\and M.\ J.\ Burgdorf \inst{\ref{hamburgmi}}
	\and S.\ Ciceri \inst{\ref{mpia}}
	\and G.\ D'Ago \inst{\ref{iiass}}
	\and R.\ Figuera Jaimes \inst{\ref{sta}, \ref{esog}}
	\and S.-H.\ Gu \inst{\ref{yunnan}, \ref{kunming}}
	\and T.\ Haugb{\o}lle \inst{\ref{nbicsp}}
	\and T.\ C.\ Hinse \inst{\ref{kasi}}
	\and D.\ Juncher \inst{\ref{nbicsp}}
	\and N.\ Kains \inst{\ref{stsci}}
	\and E.\ Kerins \inst{\ref{jodrell}}
	\and H.\ Korhonen \inst{\ref{finca}, \ref{nbicsp}}
	\and M.\ Kuffmeier \inst{\ref{nbicsp}}
	\and L.\ Mancini \inst{\ref{mpia}}
	\and N.\ Peixinho \inst{\ref{antofa}, \ref{coimbra}}
	\and A.\ Popovas \inst{\ref{nbicsp}}
	\and M.\ Rabus \inst{\ref{puc}, \ref{mpia}}
	\and S.\ Rahvar \inst{\ref{sharif}}
	\and R.\ W.\ Schmidt \inst{\ref{zfa}}
	\and C.\ Snodgrass \inst{\ref{openpss}}
	\and D.\ Starkey \inst{\ref{sta}}
	\and J.\ Surdej \inst{\ref{liege}}
	\and R.\ Tronsgaard \inst{\ref{aarhus}}
	\and C.\ von Essen \inst{\ref{aarhus}}
	\and Yi-Bo Wang \inst{\ref{yunnan}, \ref{kunming}}
	\and O.\ Wertz \inst{\ref{liege}}
	}

\institute{
	Astrophysics Group, Keele University, Staffordshire, ST5 5BG, UK \email{d.f.evans@keele.ac.uk} \label{keele}
	\and Centre for Electronic Imaging, Department of Physical Sciences, The Open University, Milton Keynes, MK7 6AA, UK \label{opencei}
	\and Niels Bohr Institute \& Centre for Star and Planet Formation, University of Copenhagen {\O}ster Voldgade 5, 1350 - Copenhagen, Denmark \label{nbicsp}
	\and SUPA, School of Physics \& Astronomy, University of St Andrews, North Haugh, St Andrews KY16 9SS, UK \label{sta}
	\and Qatar Environment and Energy Research Institute (QEERI), HBKU, Qatar Foundation, Doha, Qatar \label{qeeri}
	\and Dark Cosmology Centre, Niels Bohr Institute, University of Copenhagen, Juliane Maries Vej 30, DK-2100 Copenhagen Ø \label{nbidcc}
	\and Dipartimento di Fisica "E.R. Caianiello", Universit{\`a} di Salerno, Via Giovanni Paolo II 132, 84084, Fisciano, Italy \label{salerno}
	\and Istituto Nazionale di Fisica Nucleare, Sezione di Napoli, Napoli, Italy \label{infn}
	\and Meteorologisches Institut, Universit{\"a}t Hamburg, Bundesstra\ss{}e 55, 20146 Hamburg, Germany \label{hamburgmi}
	\and Max Planck Institute for Astronomy, K{\"o}nigstuhl 17, 69117 Heidelberg, Germany \label{mpia}
	\and Istituto Internazionale per gli Alti Studi Scientifici (IIASS), Via G. Pellegrino 19, 84019 Vietri sul Mare (SA), Italy \label{iiass}
	\and European Southern Observatory, Karl-Schwarzschild Stra\ss{}e 2, 85748 Garching bei M\"{u}nchen, Germany \label{esog}
	\and Yunnan Observatories, Chinese Academy of Sciences, Kunming 650011, China \label{yunnan}
	\and Key Laboratory for the Structure and Evolution of Celestial Objects, Chinese Academy of Sciences, Kunming 650011, China \label{kunming}
	\and Korea Astronomy \& Space Science Institute, 776 Daedukdae-ro, Yuseong-gu, 305-348 Daejeon, Republic of Korea \label{kasi}
	\and Space Telescope Science Institute, 3700 San Martin Drive, Baltimore, MD 21218, United States of America \label{stsci}
	\and Jodrell Bank Centre for Astrophysics, School of Physics and Astronomy, University of Manchester, Oxford Road, Manchester M13 9PL, UK \label{jodrell}
	\and Finnish Centre for Astronomy with ESO (FINCA), V{\"a}is{\"a}l{\"a}ntie 20, FI-21500 Piikki{\"o}, Finland \label{finca}
	\and Unidad de Astronom{\'{\i}}a, Fac. de Ciencias B{\'a}sicas, Universidad de Antofagasta, Avda. U. de Antofagasta 02800, Antofagasta, Chile \label{antofa}
	\and CITEUC -- Centre for Earth and Space Science Research of the University of Coimbra, 	Observat\'orio Astron\'omico da Universidade de Coimbra, 3030-004 Coimbra, Portugal \label{coimbra}
	\and Instituto de Astrof\'isica, Facultad de F\'isica, Pontificia Universidad Cat\'olica de Chile, Av. Vicu\~na Mackenna 4860, 7820436 Macul, Santiago, Chile \label{puc}
	\and Department of Physics, Sharif University of Technology, PO Box 11155-9161 Tehran, Iran \label{sharif}
	\and Astronomisches Rechen-Institut, Zentrum f\"ur Astronomie, Universit\"at Heidelberg, M\"onchhofstra{\ss}e 12-14, 69120 Heidelberg, Germany \label{zfa}
	\and Planetary and Space Sciences, Department of Physical Sciences, The Open University, Milton Keynes, MK7 6AA, UK \label{openpss}
	\and Institut d'Astrophysique et de G\'eophysique, All\'ee du 6 Ao\^ut 17, Sart Tilman, B\^at. B5c, 4000 Li\`ege, Belgium \label{liege}
	\and Stellar Astrophysics Centre, Department of Physics and Astronomy, Aarhus University, Ny Munkegade 120, DK-8000 Aarhus C, Denmark \label{aarhus}
	}

\date{Received -; accepted -}

\abstract{Wide binaries are a potential pathway for the formation of hot Jupiters. The binary fraction among host stars is an important discriminator between competing formation theories, but has not been well characterised. Additionally, contaminating light from unresolved stars can significantly affect the accuracy of photometric and spectroscopic measurements in studies of transiting exoplanets. }
{We observed 101 transiting exoplanet host systems in the Southern hemisphere in order to create a homogeneous catalogue of both bound companion stars and contaminating background stars, in an area of the sky where transiting exoplanetary systems have not been systematically searched for stellar companions. We investigate the binary fraction among the host stars in order to test theories for the formation of hot Jupiters.}
{Lucky imaging observations from the Two Colour Instrument on the Danish 1.54m telescope at La Silla were used to search for previously unresolved stars at small angular separations. The separations and relative magnitudes of all detected stars were measured. For 12 candidate companions to 10 host stars, previous astrometric measurements were used to evaluate how likely the companions are to be physically associated.}
{We provide measurements of 499 candidate companions within 20 arcseconds of our sample of 101 planet host stars. 51 candidates are located within 5 arcseconds of a host star, and we provide the first published measurements for 27 of these. Calibrations for the plate scale and colour performance of the Two Colour Instrument are presented.}
{We find that the overall multiplicity rate of the host stars is 38$^{+17}_{-13}$\%, consistent with the rate among solar-type stars in our sensitivity range, suggesting that planet formation does not preferentially occur in long period binaries compared to a random sample of field stars. Long period stellar companions ($P>10$ yr) appear to occur independently of short period companions, and so the population of close-in stellar companions is unconstrained by our study. }

\keywords{planets and satellites: dynamical evolution and stability -- planets and satellites: formation -- techniques: high angular resolution -- binaries: visual}

\maketitle


\section{Introduction}

\label{sec:intro}

The discovery and observation of exoplanets has posed many questions about how planets are formed. One group of planets in particular, the hot Jupiters, has been the subject of intense study -- these planets are gas giants with masses similar to Jupiter, but are found orbiting their host stars at fractions of an au, much closer than the gas giants in our own Solar System. Their orbits do not fit in with planet formation distances predicted by the core accretion model, which states that gas giants should form in the outer regions of a protoplanetary disc, with frozen volatiles being vital to their formation \citep{1996Icar..124...62P}. The inner limit for the condensation of ices in the disc is 4-5 au for Solar-type stars \citep{1995Sci...267..360B}, whereas in-situ formation of hot Jupiters would involve disc temperatures over 1,500K \citep{1996Natur.380..606L}, too hot for almost any solid material to exist.

It is now widely believed that hot Jupiters initially formed far from their host stars, as predicted by the core accretion model, and have since migrated inwards -- see the recent reviews on planet-disc interactions by \citet{2014prpl.conf..667B}, and on long term dynamical processes by \citet{2014prpl.conf..787D}. Initial work focused on interactions with the protoplanetary disc, causing the planet to lose angular momentum and spiral inwards towards the star. This migration would then be stopped by the planet reaching the inner edge of the disc, or alternatively by tidal interactions between the planet and its host star. This process would result in a well-circularised orbit with a period of only a few days \citep{1996Natur.380..606L}. However, the simple disc migration theory fails to explain the number of hot Jupiters in eccentric orbits \citep{2003ApJ...589..605W}, or those with orbits that are retrograde or misaligned compared to the rotation of their host stars \citep{2003ApJ...589..605W, 2007ApJ...669.1298F}. Whilst misaligned protoplanetary discs provide a possible pathway for this (e.g. \citealt{2010MNRAS.401.1505B}), observational studies have found that discs are generally well aligned to their host stars, as are the planets within the discs \citep{2011MNRAS.413L..71W, 2014MNRAS.438L..31G}.

Gravitational interactions with a third body can also cause changes in a planet's orbit. Outer planets can cause planet-planet scattering events \citep{1996Sci...274..954R, 2008ApJ...686..580C}, whilst inclined planetary or stellar companions can force the inner planet to undergo Kozai-Lidov oscillations, in which it is forced through alternating phases of high eccentricity and high inclination \citep{2003ApJ...589..605W, 2007ApJ...669.1298F, 2011Natur.473..187N}. These pathways are able to explain eccentric and misaligned hot Jupiters, but require a population of outer companions. Hot Jupiters rarely have nearby planetary companions \citep{2012PNAS..109.7982S}, and so if gravitational interactions are the main origin of such planets, many systems would have to be wide stellar binaries. The host stars are generally Sun-like FGK dwarfs, which have a multiplicity rate of 44$\pm$2\% \citep{2010ApJS..190....1R}, but the binary distribution among known hot Jupiter host stars is very likely to be different. Close binaries are selected against in planet-hunting surveys, due to the difficulty of detecting and characterising a planet in such a system. Additionally, there is evidence that close binaries inhibit planet formation \citep{2011A&A...528A..40F, 2012A&A...542A..92R, 2014ApJ...783....4W}. However, if gravitational interactions with a distant binary companion are a common migration pathway it would be expected that the binary fraction would be significantly enhanced.

To date, several studies have attempted to estimate the multiplicity rate among exoplanet host stars, but the results have been wildly disparate. At the low end, \citet{2012A&A...542A..92R} reported that as few as 12\% of hot Jupiters may be in multiple systems; in contrast, \citet{2015ApJ...800..138N} estimate a binary rate of 51\% from direct imaging alone, which is raised even higher when combined with radial velocity results from \citet{2014ApJ...785..126K}. However, it is difficult to compare direct imaging surveys, due to differences in sensitivity to companions and the area of the sky searched -- it would therefore be advantageous to survey or re-analyse a large number of systems in a homogeneous manner, in order to create a large sample of systems from which patterns and trends can be easily identified.

Transit searches suffer from a high rate of astrophysical false positives \citep{2003ApJ...593L.125B}, and eclipsing binary (EB) systems have proved to be a troublesome source of transit-like events, with periods and eclipse durations similar to those of hot Jupiters. The depths of the eclipses in these systems are generally much larger than would be expected by a planetary transit, but the observed depth is often reduced due to blends with nearby stars -- if an eclipsing binary is blended with another star of equal brightness, the eclipse depths will appear halved. A chance alignment of a bright foreground star and a dim background EB can cause the observed eclipses to be almost impossible to distinguish from real planetary transits, a problem that has plagued many surveys. Blending is especially problematic for surveys looking in the crowded galactic plane, such as the Lupus \citep{2009AJ....137.4368B} and OGLE \citep{2004ApJ...614..979T} collaborations, but the problem still exists in more sparsely populated fields, with the WASP-South survey finding that for every 14 candidates sent for follow-up, 13 are astrophysical mimics or blends \citep{2011EPJWC..1101004H}. The large number of planet candidates provided by the Kepler mission has resulted in several systematic searches for contaminating stars using various forms high resolution imaging, which were compared to one another by \citet{2014A&A...566A.103L}.

If a planet does indeed exist, blended light can still cause problems, due to diluted transits leading to incorrect determinations of planetary properties. An extreme case is that of Kepler-14b, where both photometric and spectroscopic measurements were affected by a companion star at 0.3\as\ separation, causing the planetary mass and radius to be underestimated by 10\% and 60\% respectively if the companion was not taken into account \citep{2011ApJS..197....3B}. A similar analysis was performed by \citet{2009A&A...498..567D} for the WASP-2, TrES-2 and TrES-4 systems, with the planetary parameters changing by up to 2$\sigma$\ when contaminating light from nearby stars was taken into account.

In this paper, we present high resolution observations of 101 southern hemisphere systems containing transiting hot Jupiters. These were used to search for nearby stars, either those physically associated with the systems, or background objects that contribute contaminating light. We also present follow-up observations of several previously discovered binary candidates, including analyses of the common proper motion of the candidates where sufficient data are available. The distribution of stars detected in our survey are compared to a statistical model in order to estimate the multiplicity rate among our targets, and this is compared to the rate among solar-type stars, and previous estimations of the multiplicity rate among planet host stars.


\section{Observations}                                                                                                               \label{sec:obs}

\begin{figure*} 
	\centering
	\includegraphics[width=17cm,angle=0]{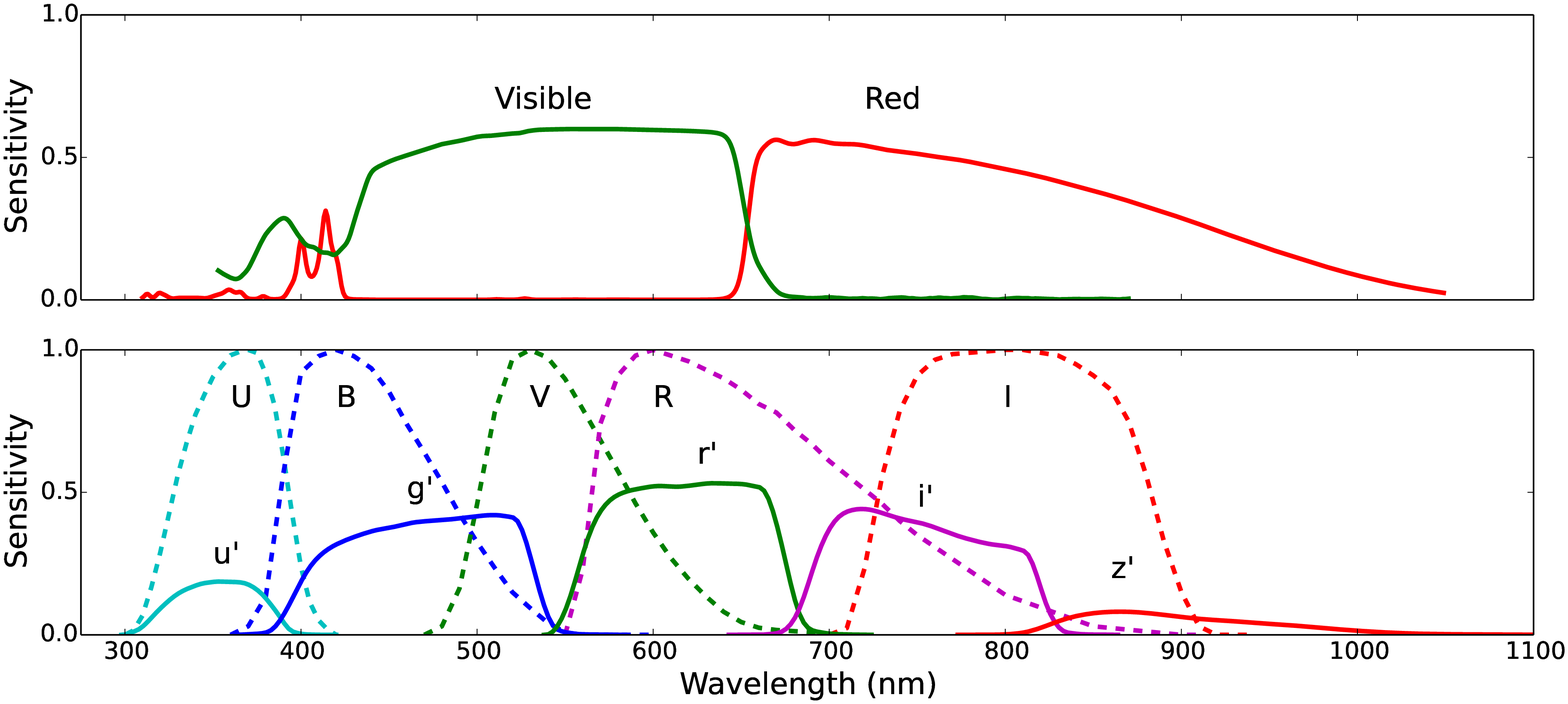}
	\caption{\label{fig:tcifilters} Top panel: The theoretical response curves of the two TCI cameras, based on the properties of the dichroics, the quantum efficiencies of the cameras, and an assumed telescope transmission efficiency of 65\%. Bottom panel: For comparison, the normalised response curves of the standard Johnson-Cousins $UBVRI$, and the measured sensitivity of the SDSS $u'g'r'i'z'$ cameras. Atmospheric effects are not included. This figure is adapted from Fig. 4 of \citet{2015A&A...574A..54S}, with the permission of the author. The SDSS curves are based	on a sensitivity determination made by J. Gunn in June 2001, available at http://www.sdss3.org/instruments/camera.php}
\end{figure*}

The observations were carried out between April and September 2014 using the Two Colour Instrument (TCI) at the Danish 1.54m telescope, La Silla, Chile. The TCI is a lucky imager designed for simultaneous two-colour photometry, using Electron Multiplying CCD (EMCCD) detectors. We give a brief summary of the instrument, with a more detailed description available in \citet{2015A&A...574A..54S}. The light arriving at the instrument is split between the two cameras using a dichroic with a cut-off wavelength of 655nm. A second dichroic sends blue light shortward of 466nm towards a focus system. The TCI is not equipped with filters, and so the light received by each camera is determined solely by the dichroics. The `visual' camera receives light between 466nm and 655nm, and the 'red' camera receives all light redward of 655nm, with the EMCCD detectors able to detect light out to approximately 1050nm. We denote the two passbands $\textrm{v}_{\textrm{\,TCI}}$ and $\textrm{r}_{\textrm{\,TCI}}$ for the visual and red cameras respectively. As shown in Fig.~\ref{fig:tcifilters}, $\textrm{r}_{\textrm{\,TCI}}$ is comparable to the combination of the SDSS $i'$ and $z'$ filters, or a Cousins $I$ filter with a wider passband, whilst $\textrm{v}_{\textrm{\,TCI}}$ has a similar central wavelength to the Johnson $V$ filter but with a significantly different response curve. For both cameras, the detector consists of a 512$\times$512 pixel array with a pixel scale of $\sim$0.09 arcsec/pixel, giving a $45\as \times 45\as$ field of view.

All target stars were observed using the red camera on the TCI. When possible, targets were also observed simultaneously with the visual camera, which was undergoing commissioning during the 2014 season. The two detectors were operated at a frame rate of 10 Hz for all observations. The use of a higher frame rate of 30 Hz was investigated, but this resulted in a poorer seeing correction. This was likely caused by the lower signal-to-noise ratio (SNR) for the shorter exposures, which resulted in the reduction pipeline being less able to select good quality frames.

Our targets were taken from the \texttt{TEPCat}\footnote{http://www.astro.keele.ac.uk/jkt/tepcat/} database of well-studied transiting extrasolar planets (TEPs) as of April 2014. All TEP systems observable from La Silla between April and September 2014 with brightnesses in the range $9\leq V\leq15$ were selected -- at the time of observing, this brightness range included all published HAT, HAT-South and WASP systems in the Southern hemisphere. We did not specifically include or exclude any systems based on the existence, or lack of, previously known companions.
	
For most targets, the default electron multiplication gain of 300 e$^{-}$/photon was used but targets brighter than $V=10.5$ required a lower gain of 100 e$^{-}$/photon, with no changes in gain being made during the observing season for a given star. A typical planetary transit results in a flux change of 1\%, which can only be mimicked by a blended eclipsing binary less than 5 magnitudes fainter than the foreground star -- a system fainter than this cannot produce an overall flux change of 1\%, even if it is completely eclipsed. To allow for such contaminating binaries to be reliably detected, the total exposure time for each target was chosen to give an SNR of 500 for a star 5 magnitudes fainter than the target in the V band, assuming that no contaminating light was diluting the signal from this star (i.e. the fainter star was well separated brighter star). The high target SNR included allowances for the rejection of a high fraction of frames in the lucky imaging process, and for shallower transit depths. In a few cases it was necessary to adjust the exposure time after an initial observation to reach the required sensitivity, with the observations being repeated -- however, the initial shorter exposures were still used in our data analysis. Variations in the total exposure time also occurred due to the automatic rejection of bad frames by the TCI pipeline. A summary of observations is given in Table~\ref{tab:obslist}, available electronically at the CDS.


\begin{table}
	\caption{\label{tab:obslist} Summary of the observations of transiting exoplanet host stars carried out. Where no exposure time is given in the `Vis.' column, the observation was carried out with in the Red band only. Variations in exposure times are mainly due to the rejection of bad frames by the TCI pipeline. This table is available in full at the CDS.}
	\centering
	\begin{tabular}{lccc} \hline \hline
		& & \multicolumn{2}{c}{Exposure Time (s)} \\
		Target & Obs. Date (BJD$_{\rm{TDB}}$) & Red & Vis. \\
		\hline
		CoRoT-1 & 2456768.9897 & 600 &  \\
		CoRoT-2 & 2456772.4179 & 440 &  \\
		CoRoT-2 & 2456783.4108 & 440 &  \\
		CoRoT-3 & 2456787.4274 & 890 &  \\
		& $\vdots$ & & \\
		CoRoT-18 & 2456924.4054 & 900 & 899 \\
		CoRoT-19 & 2456927.3924 & 900 & 899 \\
		CoRoT-20 & 2456924.3918 & 900 & 900 \\
		CoRoT-20 & 2456927.4096 & 898 & 899 \\
		& $\vdots$ & & \\
		\hline
	\end{tabular}
\end{table}


\section{Data reduction and analysis}                                                                                                \label{sec:analysis}

\subsection{TCI pipeline}

Raw lucky imaging data from the TCI are reduced automatically by the Odin pipeline, which is described fully in \citet{2015A&A...574A..54S}. The reduction pipeline performs the standard steps of bias frame removal and flat field correction, and identifies and corrects for cosmic rays. The individual exposures are then re-centred to remove shifts in the target on the detector, and ranked by quality, determined by the intensity of the central pixel relative to those surrounding it. Ten cuts in quality are made, and the exposures in each cut are stacked together and output as a ten-layer FITS cube. This allows a user to select only the best exposures for a high quality image, or instead to use more exposures and hence increase the effective exposure time. The cuts are concentrated towards the extremes of quality, and the percentage boundaries between them are: 1, 2, 5, 10, 20, 50, 90, 98, 99, 100. Therefore, the first cut contains the top 1\% of images ranked by quality, the second cut contains the next 1\% of images, and so on.

\subsection{Smearing}

In images taken with the TCI, bright stars show an apparent smearing along the image rows in the opposite direction to the EMCCD readout. It is thought that this is caused by charge transfer inefficiency \citep{2015A&A...574A..54S}, in which a small number of electrons can become trapped at the edges of the EMCCD chip during the electron multiplication phase \citep{2014JInst...9C2042B}. These electrons are released during the readout of later pixels in the same row, artificially increasing the signal in those pixels. Despite the smears being only a small fraction ($1\%$) of the total signal of their origin pixel, they are still bright enough to wash out dim stars. To correct for this effect, we assume that the smear-corrected signal $I_{x,\,y}$ at column $x$\ and row $y$ is related from the measured intensity $I^\prime_{x,\,y}$ as:
\begin{equation}
I_{x,\,y} = I^\prime_{x,\,y} - k\,\sum^{y-1}_{i=1} I_{x,\,i}
\end{equation}
where $k$ is the fraction of the charge that is trapped from one pixel and later released into another pixel. It is therefore assumed that the fraction of charge that becomes trapped per pixel is fixed, and that this charge is released at a constant rate as each subsequent pixel is read out. We also do not attempt to reassign the trapped signal back to its origin pixel, as this would have no effect on the relative brightness of stars if the same fraction of charge is lost per pixel. As the north/south diffraction spikes from the Danish telescope's secondary mirror supports are parallel to the detector readout direction, the north diffraction spike is significantly brighter than the south spike due to smearing. $k$\ was determined by assuming the north/south diffraction spikes visible on the TCI images should be of equal intensity after correction, and this constraint was best satisfied by $k=0.0000177$.


\subsection{Detection Method}

Faint stars located close to the target star run the risk of being lost in the wings of the bright target star's point spread function (PSF). Previous high resolution imaging surveys have often used some variation on PSF fitting to detect such stars (e.g. \citealt{2009A&A...498..567D, 2015ApJ...800..138N}), but this was not effective when applied to the TCI data. The Danish 1.54m Telescope suffers from triangular coma below 1\as, resulting in an asymmetric PSF that is not well matched by analytical models, and the changing observing conditions and stochastic nature of the speckles caused by seeing result in significant variations between the PSFs of stars observed even on the same night. Algorithms such as Starfinder \citep{2000A&AS..147..335D}, which derive an empirical PSF for the image, have been successfully applied to analyse TCI observations of crowded fields without requiring an input PSF model \citep{2013A&A...553A.111S, 2015A&A...573A.103S}. However, many of our observations have only a single bright star visible, and so there is a potential degeneracy between the case of a bright PSF with a nearby dim PSF, and a single bright PSF with a small aberration.

\citet{2012MNRAS.421.2498G} suggested a method involving the convolution of each image with a Gaussian of a width larger than the FWHM and then subtracting the ``blurred'' image from the original image. This is similar to the Difference of Gaussians (DoG) technique, in which an image is convolved with two Gaussians of differing widths that are then subtracted from one another, this technique having been used for purposes such as high-pass filtering and edge detection \citep{2002dip..book.....G}. To implement this, the astronomical images were treated as an ideal image of a set of point-like stars that has already been convolved with a Gaussian, due to effects such as seeing and diffraction. As the product of two Gaussians is another Gaussian, the second convolved image can be generated from the first seeing-convolved image.

For each image cube from the TCI's red camera, the first 7 quality cuts were loaded as separate images, meaning that the best 90\% of exposures were used. The 6th and 7th cuts contain the largest number of exposures, 30\% and 40\% respectively, allowing the detection of very faint stars despite the lower image quality. The final 3 cuts cover the worst 10\% of images, and have neither high resolutions nor long exposure times, and so these cuts were not used.

Each of the seven images was convolved with a Gaussian of standard deviation 4 pixels (FWHM 11.7 pixels) which was then subtracted from the original image, giving a set of difference images. In order to remove the signal from the target star, the images were divided into a series of annuli with a width of 0.5 pixels, centred on the target star -- the sub-pixel width was required to handle the steep gradient in signal around the target. Within each annulus, the mean and standard deviation of the counts per pixel were calculated. A sigma clipping algorithm was used to iteratively discount pixels with counts more than 3$\sigma$ from the mean, to avoid the mean and standard deviation becoming biased when a bright star was present. Finally, any pixels with a signal more than 1.7$\sigma$ from the annulus mean were flagged as candidate detections, with all candidates from each of the 7 cuts being compiled into a single detection list. The cut-off value of 1.7$\sigma$ was found to be the best compromise between reducing the number of real stars (as determined by eye) that were incorrectly rejected, and the increasing rate of false positives with lower cut-offs.

The combined detections for each observation were checked by eye, mainly to exclude spurious detections caused by the non-circular shape of the PSF. The results from each of our observations of a target were combined, creating a single verified list of candidate companions for each object. Measurements of the properties of the stars were done using a single stacked image using the 7 best quality cuts. Magnitudes relative to the target star were calculated using aperture photometry.

The separations and position angles for each candidate were measured by selecting a 9x9 pixel array centred on the both the target star and the companion stars, which were then super-sampled by a factor of 10. To calculate the position of a star's centre along one axis, the super-sampled image was summed parallel to the other axis, and the peak was chosen as the centre of the star. The accuracy of the position determination was tested using simulated stars, with no improvement being found beyond a super-sampling factor of 10, and was found to be accurate to within 0.5 pixels (0.044\as) for SNRs as low as 5.

For 11 candidate companions, accurate aperture photometry was not possible due to their small separation from the target star, usually combined with bad seeing conditions. Luckily, these cases all occurred on images with other well-separated stars suitable for use as PSF references, and the positions and magnitudes were measured using the PSF-fitting DAOPHOT routines in IRAF. It was assumed that the PSF shape was constant across the image, as the field of view was small enough to minimise atmospheric effects, and no variation in PSF shape was visible when images were inspected by eye.

\subsection{Astrometric calibration}

The pixel scale and detector rotation of the red camera were calibrated using observations of eight globular clusters. Using the stellar positions published in the USNO NOMAD-1 catalogue, an automatic fit to the detector scale and rotation was performed using the Starlink \texttt{AST} package and the \texttt{GAIA} image analysis tool\footnote{Starlink is available at http://starlink.eao.hawaii.edu/starlink/}. The uncertainties in the NOMAD-1 catalogue varied depending on the brightness of the cluster and sources used for the data, typically being in the range of 60 to 200 mas for each star -- however, as several dozen stars were used in each fit, the effects of random errors were reduced. Our uncertainties were derived from the scatter between the different fits generated. In order to check for any variations in scale or rotation with sky position or date, the images used for calibration were chosen to cover virtually the entire range of sky positions, and covered a range of dates from 2014-05-07 to 2014-09-21. Additionally, an earlier image from the 2013 season was used as a further check for long term consistency. 

The $+y$ axis was found to be rotated from North by $1.1\pm0.2^\circ$ with a scale of $88.7\pm0.4$ mas/pixel, and the $-x$ axis rotated from East by $1.1\pm0.3^\circ$ at a scale of $88.9\pm0.5$ mas/pixel. From these results, it was assumed that any difference in scale and rotation between the two directions was negligible, and the values were combined to give a rotation of $1.1\pm0.2^\circ$ at a scale of $88.8\pm0.3$ mas/pixel. This was not found to vary with the pointing of the telescope, and no evidence of variation with time was found.

Astrometric calibration was not performed for the visual camera, as the detector was being commissioned during the summer 2014 observing season \citep{2015A&A...574A..54S}. This resulted in the orientation of the camera changing several times. In general, the visual camera images were rotated by approximately $+$2\degrees\ compared to the red camera, and were offset slightly towards the south west.

\subsection{Colour calibration}
\label{sec:colourcalib}
In September 2014, we observed the seven spectrophotometric standard stars listed in Table~\ref{tab:standardstars} taken from the ESO Optical and UV Spectrophotometric Standard Stars catalogue\footnote{Available at http://www.eso.org/sci/observing/tools/standards.html} in order to test the colour response of the TCI. Theoretical predictions of the TCI's colour response curves were presented in \citet{2015A&A...574A..54S}, which were combined with the spectra of the standard stars to give an estimated Vis-Red colour index, abbreviated as $\left(\textrm{v}-\textrm{r}\right)_{\textrm{\,TCI}}$. The fluxes were measured using aperture photometry, and a $\left(\textrm{v}-\textrm{r}\right)_{\textrm{\,TCI}}$ colour index was generated for each observation. 

We modelled atmospheric extinction using a linear relationship, with the corrected colour $\left(\textrm{v}-\textrm{r}\right)_{\textrm{\,TCI}}$ being related to the measured colour $\left(\textrm{v}-\textrm{r}\right)_{\textrm{\,TCI}}'$ as follows:
\begin{equation}
\left(\textrm{v}-\textrm{r}\right)_{\textrm{\,TCI}} = \left(\textrm{v}-\textrm{r}\right)_{\textrm{\,TCI}}' - kZ
\end{equation}
where $Z$ is the airmass and $k$ is a coefficient fitted to the data, with a value of $+0.08\pm0.03$ mag/airmass being chosen as the best-fit value. However, there was significant scatter in the data, and improved observations would be useful in verifying and further constraining this value and its dependence on stellar colour. This agrees fairly well with the measurements of \citet{1977Msngr..11....7T} at the La Silla site, which predicts that the relative extinction coefficient of filters centred at 560nm and 860nm would be approximately $+0.1$ mag/airmass.

\begin{table} 
	\caption{\label{tab:standardstars} List of spectrophotometric stars observed.}
	\centering
	\begin{tabular}{l c c c} \hline \hline 
		Target & SpT & V magnitude & Ref. \\
		\hline
		CD-34 241 & F & 11.2 & 1 \\
		Feige-110 & DO & 11.8 & 2 \\
		G158-100 & dG & 14.9 & 3 \\
		LTT 7987 & DA & 12.2 & 2 \\
		LTT 9239 & F & 12.1 & 1 \\
		LTT 9491 & DB & 14.1 & 1 \\
		NGC 7293 & DA & 13.5 & 3 \\
		\hline
	\end{tabular}
	\tablebib{(1) \citet{1992PASP..104..533H, 1994PASP..106..566H}; (2) \citet{2014A&A...568A...9M}; (3) \citet{1990AJ.....99.1621O}.}
\end{table}

\begin{figure} 
	\includegraphics[width=\columnwidth,angle=0]{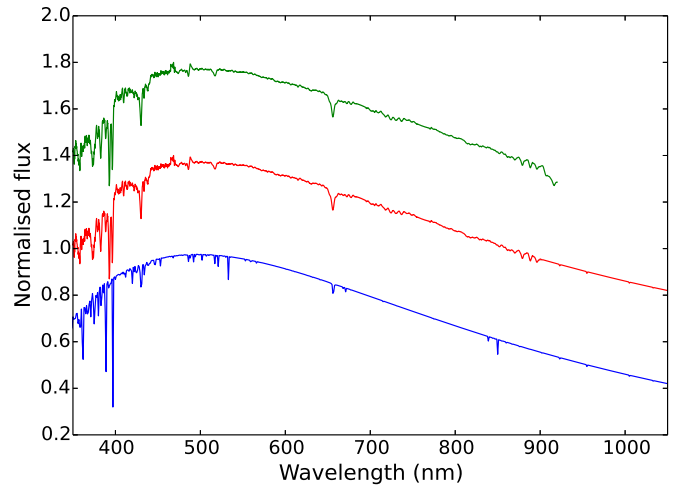}
	\caption{\label{fig:g158100} The spectra of G158-100. The top green line is the measured spectrum \citep{1990AJ.....99.1621O}, the bottom blue line the model, and the central red line the combined spectrum. All curves are scaled to a peak flux of 1.0, and are offset for clarity. } 
\end{figure}

The spectrum used for the standard star G158-100 in \citet{1990AJ.....99.1621O} is cut off redwards of 920nm, causing the flux in the red filter to be underestimated. The star is a G-type subdwarf, with \Teff$\,=\,4981$K, \logg$\,=\,4.16$, and [Fe/H]$\,=\,-2.52$ \citep{2005ApJ...633..398B}. To extend the data from 900nm onwards, we used a \texttt{PHOENIX} model spectrum \citep{2013A&A...553A...6H}, with \Teff$\,=\,5300$K, \logg$\,=\,4.00$, and [Fe/H]$\,=\,-3.0$, the temperature being varied to best fit the measured spectrum. The result is shown in Fig.~\ref{fig:g158100}.

A similar problem with cut-off spectra occurred in the cases of NGC 7293 and LTT 9491. Due to their high temperatures and relatively featureless spectra, both were fitted as black bodies. NGC 7293 was extended from 900nm onwards with a black body temperature of 110,000K, and LTT 9491 was extended from 970nm onwards with a temperature of 12,500K.

\begin{figure} 
	\includegraphics[width=\columnwidth,angle=0]{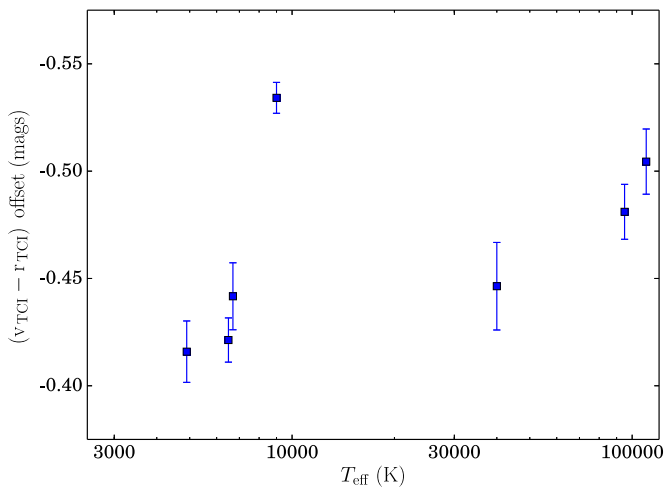}
	\caption{\label{fig:colouroffsets} The offsets between measured and predicted $\left(\textrm{v}-\textrm{r}\right)_{\textrm{\,TCI}}$ colour index for the observed standard stars, after correction for atmospheric extinction. Error bars are derived from the scatter in the measurements. } 
\end{figure}

Upon analysis of the extinction-corrected colour indices, it became apparent that all stars were systematically offset by $-0.46$ in the $\left(\textrm{v}-\textrm{r}\right)_{\textrm{\,TCI}}$ colour index compared to the predicted values, indicating that the stars appear significantly bluer than expected. This is much larger than the atmospheric extinction correction, and it is not currently clear what the cause of the offset is. There is some evidence that the colour offset is temperature related, with cooler stars generally showing a lower offset, as shown in Fig.~\ref{fig:colouroffsets}. 

\subsection{Stellar colour indices}
It is possible to measure the colour of objects by using the flux in the visual and red cameras of the TCI, and hence to estimate the effective temperature of the object. This can then be used to compare the photometric parallaxes of the target and candidate companions -- a background star will have a larger photometric parallax than the target star, hence showing that the two are not physically bound. The passbands of the two TCI cameras differ significantly from any photometric system, and no set of colour indices for standard filters matches the system well. We therefore calculated a set of theoretical colour indices for main sequence dwarf stars of solar metallicity for the EMCCD instrument, using \texttt{PHOENIX} model spectra \citep{2013A&A...553A...6H} and the passband data published in \citet{2015A&A...574A..54S}.

As the \texttt{PHOENIX} models are given in flux per unit surface area, they do not provide any information on the relative luminosity of different stellar types -- therefore, information on stellar radii was required, in order to allow relative photometric parallaxes to be determined. Stellar radii can be taken from theoretical stellar models, but these have been found to underestimate the radius of low mass stars when compared to direct observational data (e.g. \citealt{2012ApJ...746..101B, 2012ApJ...757..112B, 2015ApJ...804...64M}). Instead, it was decided to derive an empirical relation between stellar mass and effective temperature, in order to allow the colour indices to be expressed in terms of mass.

\texttt{DEBCat}\footnote{http://www.astro.keele.ac.uk/jkt/debcat/} is a catalogue of the physical properties of detached eclipsing binaries. The catalogue only includes systems in which the two stars are well separated, ensuring that their evolution has not been affected by mass transfer. The stars in the catalogue therefore provide a set of measurements that are representative of non-binary stars \citep{2015ASPC..496..164S}. However, at higher masses, the catalogue is biased to include a significant proportion of stars that have evolved off the main sequence, due to these being both brighter and more likely to eclipse than unevolved stars \citep{1991A&ARv...3...91A}. Therefore, similar cuts to those used in \citet{2009MNRAS.394..272S} were chosen, with only stars below 1.5\Msun\ included in the fit. Additionally, for systems where both components are above 1.0\Msun, only the secondary stars from systems with $q = \frac{M_B}{M_A} < 0.9$ were used. Assuming that both stars in a binary system are co-evolutionary, a higher mass primary star will evolve off the main sequence first. By choosing only secondary stars that were significantly less massive than the primary stars, the difference in the two components' main sequence lifetimes ensured that even if the primary had begun to evolve, the secondary would still be on the main sequence. For systems where both components were below 1.0\Msun, the main sequence lifetimes of both components were assumed to be long enough that neither star was likely to have evolved significantly. The V1174 Ori system was manually excluded despite matching these criteria, as it comprises two pre-main sequence stars \citep{2004ApJS..151..357S}, which have much larger radii than main sequence stars.

\begin{figure} 
	\includegraphics[width=\columnwidth,angle=0]{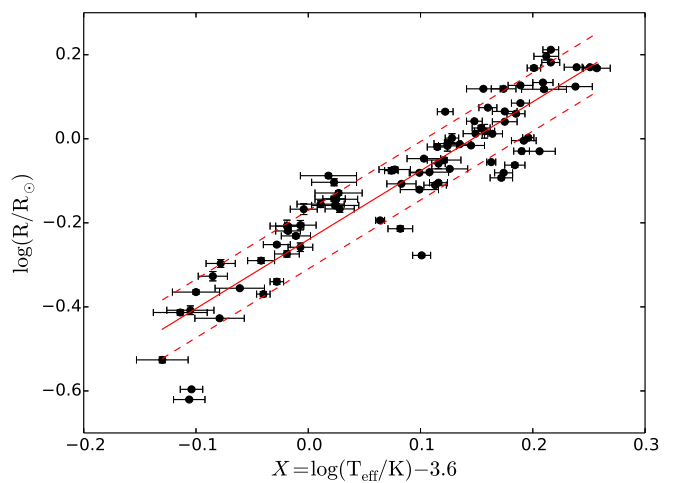}
	\caption{\label{fig:tempradiusfit} The empirical relationship between stellar radius and effective temperature, derived from the sample of detached eclipsing binaries. The dashed lines indicate the RMS scatter. } 
\end{figure}

\begin{figure} 
	\includegraphics[width=\columnwidth,angle=0]{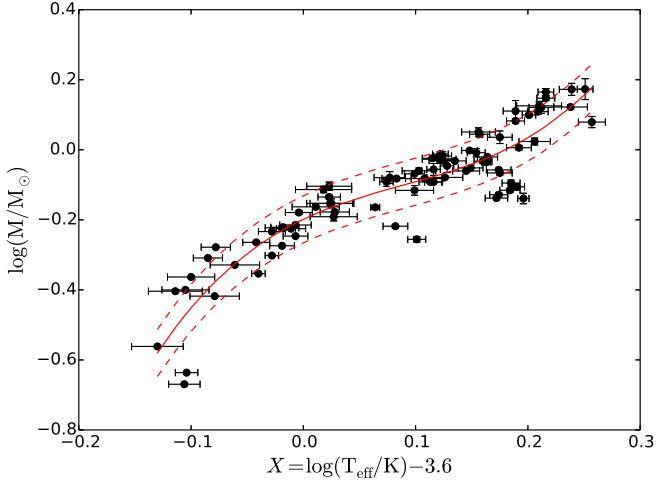}
	\caption{\label{fig:tempmassfit} The empirical relationship between stellar mass and effective temperature, derived from the sample of detached eclipsing binaries. The dashed lines indicate the RMS scatter. } 
\end{figure}

Rather than fitting directly to the effective temperature of the star, \Teff, we define the variable $X=\log(\Teff)-3.6$, to increase the numerical accuracy of our fits. The linear fit between $X$ and $\log\,(\mathrm{R}\,/\Rsun)$ is shown in Fig.~\ref{fig:tempradiusfit}, and the cubic fit between $X$ and $\log\,(\mathrm{M}\,/\Msun)$ in Fig.~\ref{fig:tempmassfit}. The scatter in the data points is larger than would be expected from the error bars alone, and can be attributed to physical differences between the stars in the dataset, caused by properties such as metallicity or stellar activity which were not considered in this fit. Because of this, the fit was not improved significantly by the inclusion of measurement errors. Instead, we quote the RMS scatter.

The radius-temperature relation derived is,
\begin{equation}
\log\,(\mathrm{R}/\Rsun) = 1.639\cdot X - 0.240
\end{equation}
with $R$ being the stellar radius, and an RMS scatter of 0.07 around the fit. The mass-temperature relation is,
\begin{eqnarray}
\log\,(\mathrm{M}/\Msun) = 27.296\cdot X^3 - 7.273\cdot X^2 \nonumber \\
+ 1.529\cdot X - 0.198
\end{eqnarray}
with $M$ being the stellar mass, the data having an RMS scatter of 0.08 around the fit.

The derived colour indices are presented in Table~\ref{tab:colourindices}. The values of surface gravity used for the model atmospheres are listed in the table, and were chosen based on the values given in \citet{1976oasp.book.....G}. Note that these indices are not corrected for the systematic colour offset discussed in Section~\ref{sec:colourcalib}.

\begin{table} \centering \caption{\label{tab:colourindices} 
		\reff{Theoretical colour indices and relative magnitudes for the EMCCD instrument. log g is the logarithm of the stellar surface gravity, in cgs units. $\Delta \textrm{r}_{\textrm{\,TCI}}$ is the difference in $\textrm{r}_{\textrm{\,TCI}}$ magnitude compared to a star with \Teff$=$7000K, and $\Delta \textrm{v}_{\textrm{\,TCI}}$ is the same for the $\textrm{v}_{\textrm{\,TCI}}$ magnitude.}}
	\begin{tabular}{c c c c c c} \hline \hline
		\Teff\ (K) & M ($\Msun$) & log g & $ \left(\textrm{v}-\textrm{r}\right)_{\textrm{\,TCI}}$ & $\Delta \textrm{r}_{\textrm{\,TCI}}$ & $\Delta \textrm{v}_{\textrm{\,TCI}}$\\
		\hline
		12000 & & 4.0 & $-$0.50 & & \\ 
		11600 & & 4.0 & $-$0.50 & & \\ 
		11200 & & 4.0 & $-$0.49 & & \\ 
		10800 & & 4.0 & $-$0.48 & & \\ 
		10400 & & 4.0 & $-$0.47 & & \\ 
		10000 & & 4.0 & $-$0.45 & & \\ 
		9600 & & 4.0 & $-$0.44 & & \\ 
		9200 & & 4.0 & $-$0.42 & & \\ 
		8800 & & 4.5 & $-$0.38 & & \\ 
		8400 & & 4.5 & $-$0.34 & & \\ 
		8000 & & 4.5 & $-$0.29 & & \\ 
		7600 & & 4.5 & $-$0.23 & & \\ 
		7200 & & 4.5 & $-$0.16 & & \\ 
		7000 & 1.38 & 4.5 & $-$0.13 & 0.0 & 0.0 \\
		6800 & 1.28 & 4.5 & $-$0.09 & 0.2 & 0.3 \\
		6600 & 1.19 & 4.5 & $-$0.06 & 0.4 & 0.5 \\
		6400 & 1.11 & 4.5 & $-$0.02 & 0.7 & 0.8 \\
		6200 & 1.05 & 4.5 & 0.02 & 0.9 & 1.0 \\
		6000 & 0.99 & 4.5 & 0.06 & 1.1 & 1.3 \\
		5800 & 0.95 & 4.5 & 0.10 & 1.4 & 1.6 \\
		5600 & 0.91 & 4.5 & 0.15 & 1.6 & 1.9 \\
		5400 & 0.87 & 4.5 & 0.20 & 1.9 & 2.2 \\
		5200 & 0.84 & 4.5 & 0.25 & 2.2 & 2.6 \\
		5000 & 0.81 & 4.5 & 0.32 & 2.5 & 3.0 \\
		4900 & 0.79 & 4.5 & 0.35 & 2.7 & 3.1 \\
		4800 & 0.78 & 4.5 & 0.39 & 2.8 & 3.4 \\
		4700 & 0.77 & 4.5 & 0.43 & 3.0 & 3.6 \\
		4600 & 0.75 & 4.5 & 0.48 & 3.2 & 3.8 \\
		4500 & 0.74 & 4.5 & 0.53 & 3.4 & 4.0 \\
		4400 & 0.72 & 4.5 & 0.58 & 3.6 & 4.3 \\
		4300 & 0.70 & 4.5 & 0.65 & 3.8 & 4.6 \\
		4200 & 0.68 & 4.5 & 0.71 & 4.0 & 4.8 \\
		4100 & 0.66 & 4.5 & 0.79 & 4.2 & 5.1 \\
		4000 & 0.64 & 4.5 & 0.86 & 4.4 & 5.4 \\
		3900 & 0.61 & 5.0 & 0.94 & 4.6 & 5.7 \\
		3800 & 0.59 & 5.0 & 1.03 & 4.8 & 6.0 \\
		3700 & 0.56 & 5.0 & 1.13 & 5.1 & 6.3 \\
		3600 & 0.52 & 5.0 & 1.24 & 5.3 & 6.7 \\
		3500 & 0.49 & 5.0 & 1.36 & 5.6 & 7.1 \\
		3400 & 0.45 & 5.0 & 1.49 & 5.9 & 7.5 \\
		3300 & 0.41 & 5.0 & 1.63 & 6.2 & 7.9 \\
		3200 & 0.37 & 5.0 & 1.80 & 6.5 & 8.4 \\
		3100 & 0.33 & 5.0 & 1.99 & 6.8 & 8.9 \\
		3000 & 0.28 & 5.0 & 2.21 & 7.1 & 9.5 \\
		2900 & 0.24 & 5.0 & 2.47 & 7.5 & 9.7 \\
		2800 & 0.20 & 5.0 & 2.77 & 7.9 & 10.8 \\
		2700 & 0.16 & 5.0 & 3.11 & 8.3 & 11.6 \\
		2600 & 0.12 & 5.0 & 3.39 & 8.8 & 12.3 \\
		2500 & 0.09 & 5.0 & 3.57 & 9.2 & 12.9 \\
		2400 & 0.07 & 5.0 & 3.51 & 9.5 & 13.1 \\
		2300 & 0.05 & 5.0 & 3.48 & 9.9 & 13.5 \\
		
		\hline \end{tabular}
\end{table}

\subsection{Host star distances} \label{sec:distance}

To convert between the physical and projected separations of two stars, their distance must be known. A number of different methods have been used in the literature to calculate the distances to TEP host stars, and it is not always clear which method was used. We therefore decided to derive a homogeneous set of distances to the targeted host stars using the K-band surface brightness-effective temperature relation presented in \citet{2004A&A...426..297K}.

The stellar radius and effective temperature were taken from the \texttt{TEPCat} database of physical properties of transiting exoplanet systems \citep{2011MNRAS.417.2166S}, using the data available as of 2015-10-01\footnote{\texttt{TEPCat} is archived monthly, and the version used in this work can be found at: http://www.astro.keele.ac.uk/jkt/tepcat/2015oct/allplanets-noerr.html}, with the K band magnitudes derived from \texttt{2MASS} Ks magnitudes \citep{2006AJ....131.1163S}. Stars separated by less than 5\as\ were not consistently detected as two separate stars by the \texttt{2MASS} data reduction pipeline, and so the K band magnitudes suffer from contaminating light. However, in all cases we find that the candidate companions are faint enough that they contribute only a few percent of the total flux, and are often sufficiently separated for contaminating light to be at a relatively low level. K band extinction was assumed to be negligible for the entire sample. Our derived distances are listed in Table~\ref{tab:distances}, available electronically from the CDS, and are generally within $2\sigma$ of the previously reported values. Notably different distances are those for WASP-66 and WASP-67, for which we find $520\pm30$pc and $184\pm6$pc respectively, compared to $380\pm100$ and $255\pm45$ given by \citet{2012MNRAS.426..739H}; WASP-90, which we find to be at $211\pm10$ compared to $340\pm60$ by \citet{2016A&A...585A.126W}; and OGLE-TR-113, which we place at $358\pm19$pc, much closer than the previously reported distances of $600$pc by \citet{2004ApJ...609L..37K} and $553$pc by \citet{2006A&A...458..997S}.

\begin{table} \centering \caption{\label{tab:distances} 
		\reff{Distances derived for planet host stars using the K-band surface brightness-effective temperature relation. This table is available in full at the CDS.}}
	\begin{tabular}{c l} \hline \hline
		System & Distance (pc) \\
		\hline
		    CoRoT-1 & $ 783\pm 36$ \\
		    CoRoT-2 & $ 226\pm  6$ \\
			CoRoT-3 & $ 800\pm 50$ \\
			& $\vdots$ \\
		\hline \end{tabular}
\end{table}

\section{Results}                                                                                                               \label{sec:results}



We report 25 new candidate companions within 5 arcseconds of 18 planet host stars. We also present new observations of 14 companions within 5 arcseconds of 11 planet host stars. Candidate companions to CoRoT stars are presented in Table~\ref{tab:closecomps_CoRoT_combined}, HAT, HAT-South and WASP stars in Table~\ref{tab:closecomps_HATWASP_combined}, and OGLE stars in Table~\ref{tab:closecomps_OGLE_combined}.

In the sections below we discuss close candidate companions within 5\as\ where data has allowed us to make further conclusions, including twelve systems where either previous astrometric measurements have allowed us to study the candidate's relative proper motion, and four systems where two colour imaging with the TCI has permitted us to compare the photometric distances to the two stars.

A full list of all stars found within 20\as\ of a TEP host star, along with their measured positions and relative brightnesses for each observation, can be found in Table~\ref{tab:complist}, available electronically from the CDS.

\subsection{Common proper motion analysis}

If two stars are gravitationally bound to one another, it is expected that they will display common proper motion. Therefore, the separation and position angle of the companion relative to the planet host star should not change significantly with time, assuming that orbital motion is negligible over the observational baseline. An alternative scenario is that in which the candidate companion is a distant background star, showing negligible proper motion compared to the foreground target star. In this case, it is expected that the foreground star will move past the background star over time. Additionally, the apparent separation and angle between the two stars will vary over the Earth's orbit due to parallax, with the foreground star having a significantly larger parallactic motion.

For proper motions, the UCAC4 catalogue \citep{2013AJ....145...44Z} was chosen as the source for all targets. We compared the data in UCAC4 with that in the NOMAD \citep{2004AAS...205.4815Z} and PPMXL \citep{2010AJ....139.2440R} catalogues. We generally found that all three catalogues agreed on the proper motions to within the errors, although differences were found for CoRoT-3, CoRoT-11, and HAT-P-41 -- it is likely that these discrepancies are caused by contamination from the stars detected in this survey, combined with low proper motions. We also investigated the proper motions available in the recently published URAT1 catalogue \citep{2015AJ....150..101Z}. We found that the values were often significantly different to the other three catalogues for our targets, even those with high proper motions, and therefore decided not to use the URAT1 data.

We generated a model describing the relative motion of the two objects assuming that the candidate companion was a distant background star. The expected motion was then calculated using the position of the companion, which was allowed to vary to fit the data, and the UCAC4 proper motion of the target and the parallax expected from the distances derived in Section~\ref{sec:distance}. We also created a second model in which the two stars showed no relative motion, and again fitted the initial position of the companion to give the best fit. The two models were compared using the $\chi^2$ goodness of fit, from which the more probable model was identified, and the significance with which it was preferred. The fit comparisons are shown in Table~\ref{tab:propermotioncomparison}. Figs.~\ref{fig:cpm1} and \ref{fig:cpm2} show the measured separations and position measurements of twelve candidate companions for which other observations are available, as well as the best fits from the proper motion models.


\begin{table*}
	\centering
	\caption{\label{tab:closecomps_CoRoT_combined} Our measurements of candidate companions to CoRoT planet host stars. `N' indicates the number of observations used for each star. Where more than one observation was used, the MJD quoted is in the middle of the range of observing dates. The `Bound?' column indicates cases where our analyses determined that the companion is bound or unbound, and the method with which this was confirmed (Proper Motion, M, Relative Brightness, B, or Stellar Colours, C) -- no entry is given if our analyses were inconclusive. The references column lists previous analyses (if any) of the candidates. }
	\begin{tabular}{l c c c c c c l l }
		\hline \hline
		Target & MJD & N & Separation (\as) & Pos. angle (\degrees) & $\Delta \textrm{r}_{\textrm{\,TCI}}$ & $\Delta \textrm{v}_{\textrm{\,TCI}}$ & Bound? & References \\
		\hline
		CoRoT-2 & 56778 & 2 & 4.061$\pm$0.012 & 208.40$\pm$0.15 & 2.901$\pm$0.002 & -- & -- & 1, 5, 6, 7, 8\\
		CoRoT-3 (1) & 56787 & 1 & 5.19$\pm$0.02 & 174.7$\pm$0.2 & 2.589$\pm$0.004 & -- & Yes (M) & 2, 6, 7, 8\\
		CoRoT-3 (2) & 56787 & 1 & 4.96$\pm$0.02 & 91.6$\pm$0.2 & 4.8$\pm$0.3 & -- & -- & 2, 7 \\
		CoRoT-3 (3) & 56787 & 1 & 2.501$\pm$0.013 & 135.8$\pm$0.2 & 6.295$\pm$0.019 & -- & -- & -- \\
		CoRoT-3 (4) & 56787 & 1 & 3.90$\pm$0.03 & 240.7$\pm$0.3 & 8.0$\pm$0.6 & -- & -- & --\\
		CoRoT-3 (5) & 56787 & 1 & 4.51$\pm$0.19 & 130.3$\pm$0.2 & 7.6$\pm$0.4 & -- & -- & -- \\
		CoRoT-4 (1) & 56769 & 1 & 2.935$\pm$0.014 & 336.8$\pm$0.2 & 5.231$\pm$0.014 & -- & -- & -- \\
		CoRoT-4 (2) & 56769 & 1 & 4.187$\pm$0.018 & 203.2$\pm$0.2 & 6.9$\pm$0.3 & -- & -- & -- \\
		CoRoT-6 & 56785 & 2 & 4.620$\pm$0.013 & 262.53$\pm$0.15 & 6.63$\pm$0.19 & -- & -- & -- \\
		CoRoT-7 & 56769 & 1 & 4.98$\pm$0.02 & 160.4$\pm$0.2 & 8.9$\pm$0.4 & -- & No (B) & 3 \\
		CoRoT-8 (1) & 56840 & 4 & 2.468$\pm$0.009 & 39.53$\pm$0.12 & 7.174$\pm$0.024 & Undetected & -- & -- \\
		CoRoT-8 (2) & 56840 & 4 & 3.163$\pm$0.007 & 342.44$\pm$0.11 & 6.468$\pm$0.019 & Undetected & -- & -- \\
		CoRoT-8 (3) & 56840 & 4 & 4.290$\pm$0.009 & 23.78$\pm$0.11 & 6.236$\pm$0.011 & 7.0$\pm$0.6 & No (C) & -- \\
		CoRoT-9 (1) & 56783 & 1 & 4.44$\pm$0.02 & 271.2$\pm$0.2 & 7.10$\pm$0.05 & -- & -- & -- \\
		CoRoT-9 (2) & 56783 & 1 & 4.618$\pm$0.019 & 258.4$\pm$0.2 & 5.185$\pm$0.013 & -- & -- & -- \\
		CoRoT-11 (1) & 56778 & 2 & 2.537$\pm$0.009 & 307.01$\pm$0.15 & 1.994$\pm$0.002 & -- & -- & 4, 7, 8\\
		CoRoT-11 (2) & 56778 & 2 & 3.043$\pm$0.011 & 162.95$\pm$0.15 & 7.228$\pm$0.028 & -- & -- & -- \\
		CoRoT-11 (3) & 56778 & 2 & 4.006$\pm$0.014 & 222.27$\pm$0.17 & 6.81$\pm$0.18 & -- & -- & -- \\
		\hline
	\end{tabular}
	\tablebib{
		(1) \citet{2008A&A...482L..21A}; (2) \citet{2008A&A...491..889D}; (3) \citet{2009A&A...506..287L}; (4) \citet{2010A&A...524A..55G}; (5) \citet{2011A&A...532A...3S}; (6) \citet{2013MNRAS.433.2097F}; (7) \citet{2015A&A...575A..23W}; (8) \citet{2015A&A...579A.129W}.
	}
\end{table*}

\begin{table*}
	\centering
	\caption{\label{tab:closecomps_HATWASP_combined} Our measurements of candidate companions to HAT, HATS, and WASP planet host stars. `N' indicates the number of observations used for each star. Where more than one observation was used, the MJD quoted is in the middle of the range of observing dates. The `Bound?' column indicates cases where our analyses determined that the companion is bound or unbound, and the method with which this was confirmed (Proper Motion, M, Relative Brightness, B, or Stellar Colours, C) -- no entry is given if our analyses were inconclusive. The references column lists previous analyses (if any) of the candidates.}
	\begin{tabular}{l c c c c c c l l }
		\hline \hline
		Target & MJD & N$_{obs}$ & Separation (\as) & Pos. angle (\degrees) & $\Delta \textrm{r}_{\textrm{\,TCI}}$ & $\Delta \textrm{v}_{\textrm{\,TCI}}$ & Bound? & References \\
		\hline
		HAT-P-30 & 56770 & 1 & 3.856$\pm$0.016 & 3.9$\pm$0.2 & 4.333$\pm$0.017 & -- & Yes (M) & 4 9 10 12 \\
		HAT-P-35 & 56769 & 1 & 1.016$\pm$0.011 & 149.4$\pm$0.2 & 3.81$\pm$0.13 & -- & -- & 14 \\
		HAT-P-41 (1) & 56768 & 1 & 3.599$\pm$0.016 & 183.7$\pm$0.2 & 3.425$\pm$0.004 & -- & -- & 6 13 14 \\
		HAT-P-41 (2) & 56768 & 1 & 0.987$\pm$0.011 & 189.8$\pm$0.2 & 4.42$\pm$0.09 & -- & -- & -- \\
		HATS-2 & 56779 & 4 & 1.022$\pm$0.005 & 42.76$\pm$0.09 & 3.93$\pm$0.03 & -- & -- & -- \\
		HATS-3 & 56768 & 1 & 3.671$\pm$0.016 & 108.7$\pm$0.21 & 7.59$\pm$0.15 & -- & -- & -- \\  
		WASP-2 & 56772 & 1 & 0.709$\pm$0.010 & 103.4$\pm$0.2 & 4.95$\pm$0.15 & -- & Yes (M) & 1 2 8 9 12 13\\
		WASP-7 & 56779 & 3 & 4.414$\pm$0.011 & 228.73$\pm$0.12 & 9.38$\pm$0.02 & -- & -- & -- \\
		WASP-8 & 56883 & 7 & 4.495$\pm$0.007 & 170.93$\pm$0.08 & 3.534$\pm$0.002 & 4.486$\pm$0.006 & Yes (M) & 3 12 \\
		WASP-36 & 56770 & 1 & 4.872$\pm$0.019 & 66.5$\pm$0.2 & 4.579$\pm$0.018 & -- & -- & 5, 14 \\
		WASP-45 & 56847 & 1 & 4.364$\pm$0.018 & 317.7$\pm$0.2 & 6.381$\pm$0.011 & -- & -- & -- \\
		WASP-49 & 56769 & 1 & 2.239$\pm$0.013 & 178.2$\pm$0.2 & 4.979$\pm$0.018 & -- & -- & -- \\
		WASP-55 & 56783 & 3 & 4.345$\pm$0.010 & 163.62$\pm$0.12 & 5.210$\pm$0.018 & -- & -- & -- \\
		WASP-64 & 56769 & 2 & 4.55$\pm$0.02 & 324.1$\pm$0.2 & 8.3$\pm$0.4 & -- & -- & -- \\
		WASP-67 & 56768 & 1 & 4.422$\pm$0.018 & 51.6$\pm$0.2 & 7.23$\pm$0.10 & -- & -- & -- \\
		WASP-70 & 56768 & 1 & 3.140$\pm$0.015 & 167.0$\pm$0.2 & 2.1507$\pm$0.0009 & -- & Yes (M) & 11 14 \\
		WASP-77 & 56883 & 9 & 3.274$\pm$0.005 & 153.68$\pm$0.07 & 1.512$\pm$0.002 & 1.698$\pm$0.004 & Yes (MC) & 7 14 \\
		WASP-90 & 56786 & 1 & 0.992$\pm$0.011 & 184.9$\pm$0.2 & 3.22$\pm$0.09 & -- & -- & -- \\
		WASP-100 & 56921 & 2 & 3.982$\pm$0.012 & 186.30$\pm$0.15 & 6.152$\pm$0.008 & 6.18$\pm$0.05 & No (C) & -- \\
		\hline
		
	\end{tabular}
	\tablebib{
		(1) \citet{2007MNRAS.375..951C}; (2) \citet{2009A&A...498..567D}; (3) \citet{2010A&A...517L...1Q}; (4) \citet{2011AJ....142...86E}; (5) \citet{2012AJ....143...81S}; (6) \citet{2012AJ....144..139H}; (7) \citet{2013PASP..125...48M}; (8) \citet{2013MNRAS.428..182B}; (9) \citet{2013AJ....146....9A}; (10) \citet{2013A&A...559A..71G}; (11) \citet{2014MNRAS.445.1114A}; (12) \citet{2015ApJ...800..138N}; (13) \citet{2015A&A...575A..23W}; (14) \citet{2015A&A...579A.129W}.
	}
\end{table*}

\begin{table*}
	\centering
	\caption{\label{tab:closecomps_OGLE_combined} Our measurements of candidate companions to OGLE planet host stars. `N' indicates the number of observations used for each star. Where more than one observation was used, the MJD quoted is in the middle of the range of observing dates. No previous analyses have been published for the stars in this table.}
	\begin{tabular}{l c c c c c c }
		\hline \hline
		Target & MJD & N$_{obs}$ & Separation (\as) & Pos. angle (\degrees) & $\Delta \textrm{r}_{\textrm{\,TCI}}$ & $\Delta \textrm{v}_{\textrm{\,TCI}}$ \\
		\hline
		OGLE-TR-113 (1) & 58785 & 4 & 2.534$\pm$0.007 & 19.38$\pm$0.11 & 2.94$\pm$0.02 & -- \\
		OGLE-TR-113 (2) & 58785 & 4 & 3.015$\pm$0.008 & 343.42$\pm$0.11 & 4.73$\pm$0.14 & -- \\
		OGLE-TR-113 (3) & 58785 & 4 & 3.255$\pm$0.007 & 179.81$\pm$0.11 & -0.546$\pm$0.002 & -- \\
		OGLE-TR-113 (4) & 58785 & 4 & 4.27$\pm$0.04 & 23.2$\pm$0.4 & 6.4$\pm$0.5 & -- \\
		OGLE-TR-113 (5) & 58785 & 4 & 4.882$\pm$0.010 & 133.65$\pm$0.11 & 4.62$\pm$0.10 & -- \\
		OGLE-TR-113 (6) & 58785 & 4 & 4.90$\pm$0.05 & 229.1$\pm$0.5 & 6.6$\pm$0.6 & -- \\
		OGLE-TR-113 (7) & 58785 & 4 & 4.934$\pm$0.010 & 267.08$\pm$0.11 & 1.901$\pm$0.008 & -- \\
		OGLE-TR-211 & 56784 & 3 & 4.222$\pm$0.010 & 25.49$\pm$0.12 & 3.944$\pm$0.04 & -- \\
		OGLE2-TR-L9 (1) & 56786 & 5 & 2.331$\pm$0.006 & 140.66$\pm$0.09 & 5.193$\pm$0.009 & -- \\
		OGLE2-TR-L9 (2) & 56786 & 5 & 2.486$\pm$0.006 & 261.49$\pm$0.09 & 5.203$\pm$0.009 & -- \\
		OGLE2-TR-L9 (3) & 56786 & 5 & 4.084$\pm$0.008 & 106.40$\pm$0.09 & 5.182$\pm$0.009 & -- \\
		OGLE2-TR-L9 (4) & 56786 & 5 & 4.24$\pm$0.11 & 191.4$\pm$1.2 & 7.9$\pm$1.0 & -- \\
		OGLE2-TR-L9 (5) & 56786 & 5 & 4.540$\pm$0.013 & 341.50$\pm$0.13 & 6.5$\pm$0.3 & -- \\
		OGLE2-TR-L9 (6) & 56786 & 5 & 4.552$\pm$0.10 & 114.61$\pm$0.10 & 6.22$\pm$0.19 & -- \\
		\hline		
	\end{tabular}
\end{table*}

\begin{table*}
	\caption{\label{tab:complist} Observations and basic properties of all detected stars. This table is available in full at the CDS.}
	\centering
	\begin{tabular}{llccccl} \hline \hline
		& & \multicolumn{2}{c}{Exposure Time (s)} \\
		Target & ID & Separation (\as) & Position angle (\degrees) & $\Delta \textrm{r}_{\textrm{\,TCI}}$ & $\Delta \textrm{v}_{\textrm{\,TCI}}$ & Obs. Date (BJD$_{\rm TDB}$) \\
		\hline
		CoRoT-1 & 1 & 8.67$\pm$0.03 & 227.9$\pm$0.2 & 4.46$\pm$0.03 & & 2456769.4897 \\
		CoRoT-1 & 2 & 9.62$\pm$0.15 & 121.5$\pm$1.6 & 8.7$\pm$1.5 & & 2456769.4897 \\
		CoRoT-1 & 3 & 11.51$\pm$0.04 & 178.1$\pm$0.2 & 3.221$\pm$0.010 & & 2456769.4897 \\
		& & & $\vdots$ & & & \\
		CoRoT-20 & 1 & 6.46$\pm$0.02 & 76.9$\pm$0.2 & 5.7$\pm$0.3 & 7.6$\pm$1.2 & 2456924.8918 \\
		CoRoT-20 & 2 & 9.96$\pm$0.04 & 91.5$\pm$0.3 & 6.7$\pm$0.6 &  & 2456924.8918 \\
		CoRoT-20 & 3 & 10.54$\pm$0.04 & 265.7$\pm$0.2 & 5.35$\pm$0.18 & 6.6$\pm$0.5 & 2456924.8918 \\
		& & & $\vdots$ & & & \\		
		\hline
	\end{tabular}
\end{table*}


\subsection{CoRoT-2}

A star located 4\as\ from CoRoT-2 was discovered by \citet{2008A&A...482L..21A}, with infrared photometry indicating that the companion was consistent with being a physically associated K/M dwarf. \citet{2011A&A...532A...3S} obtained separate spectra of the two stars in 2010, with the companion's spectral type determined to be K9V, and measurements of the radial velocities of the two objects were consistent with a bound orbit. Further lucky imaging observations have since been published, which support the companion being in the late K/early M regime \citep{2013MNRAS.433.2097F, 2015A&A...575A..23W, 2015A&A...579A.129W}. We present two new observations of the companion shown in  Fig.~\ref{fig:cpm1}, but are unable to distinguish between the common proper motion and background star hypotheses.


\subsection{CoRoT-3}

Two nearby bright stars to CoRoT-3 were reported in \citet{2008A&A...491..889D}, both at a separation of approximately 5\as. The stars have become known as the `South' companion, 2.9\,mag fainter and denoted `1' in our work, and the `East' companion, 4.9\,mag fainter and denoted `2'. Observations of both stars were presented in \citet{2015A&A...575A..23W}, and companion 1 has been observed a further two times \citep{2013MNRAS.433.2097F, 2015A&A...579A.129W}. Whilst companion `1' is slightly beyond 5\as, we include it in this section due to its brightness.

We report one new observation of both previously reported companions. We note that the UCAC4, PPMXL, and NOMAD catalogues give significantly different values for the proper motion of CoRoT-3, and it was decided to test the motion of both companions against each of the three sets of data. We find that for the values from the UCAC4 catalogue, companion 1 is compatible with both the common proper motion and background star hypotheses. However, common proper motion is preferred when using data from the PPMXL catalogue ($2.3\sigma$) and the NOMAD catalogue ($4.3\sigma$). The situation for companion 2 is currently inconclusive for all sets of proper motion values. The position measurements for both of these companions are illustrated in Fig.~\ref{fig:cpm1}.

We also announce three new nearby stars. Star 3 is 2.5\as\ to the southeast, 6.3\,mag fainter than CoRoT-3. Stars 4 and 5 are both around 8 magnitudes fainter, at separations of 3.9\as\ and 4.5\as\ respectively.

\subsection{CoRoT-7}
\citet{2009A&A...506..287L} presented observations of 3 faint, red stars near to CoRoT-7, the nearest being located at 4.9\as\ and all being ${\sim}8.2$ mag fainter in the $J$ band. We observed all three stars again, with the two stars at larger separation being beyond the 5\as\ limit and so excluded from Table~\ref{tab:closecomps_CoRoT_combined}. All three stars are 9\,mag fainter in $\textrm{r}_{\textrm{\,TCI}}$. In both $J$ and $\textrm{r}_{\textrm{\,TCI}}$, the stars are too faint to be bound main sequence stars.

\subsection{CoRoT-8}
Of the three detected candidate companions, only companion 3 was bright enough to be detected in $\textrm{v}_{\textrm{\,TCI}}$. From its estimated effective temperature of 3420$\pm$130K, we conclude that it is four mags too faint to be bound. 

\subsection{CoRoT-11}
A nearby bright companion to CoRoT-11 was noted in \citet{2010A&A...524A..55G}, 2.1\,mag fainter in $R$ at a separation of 2\as, and was further observed using lucky imaging in 2013 and 2014 by \citet{2015A&A...575A..23W} and \citet{2015A&A...579A.129W}. We reobserved the companion, which we denote `1', and find a position consistent with previous measurements. The NOMAD catalogue gives a much larger proper motion for CoRoT-11 than the UCAC4 or PPMXL, and whilst a companion with common proper motion is preferred by the NOMAD data, the motions in the other catalogues only slightly prefer this scenario. The motion predicted by the UCAC4 catalogue is shown in Fig.~\ref{fig:cpm1}.

We also report two new faint stars, with companion 2 being 7.2\,mag fainter at a separation of 3\as, and companion 3 being 6.8\,mag fainter at 4\as. Due to CoRoT-11's high effective temperature of 6440$\pm$120K \citep{2010A&A...524A..55G}, it is possible that these could be faint bound M dwarfs.

\subsection{HAT-P-30}
\begin{figure*} 
	\centering
	\begin{tabular}{cc}
		\includegraphics[width=8cm]{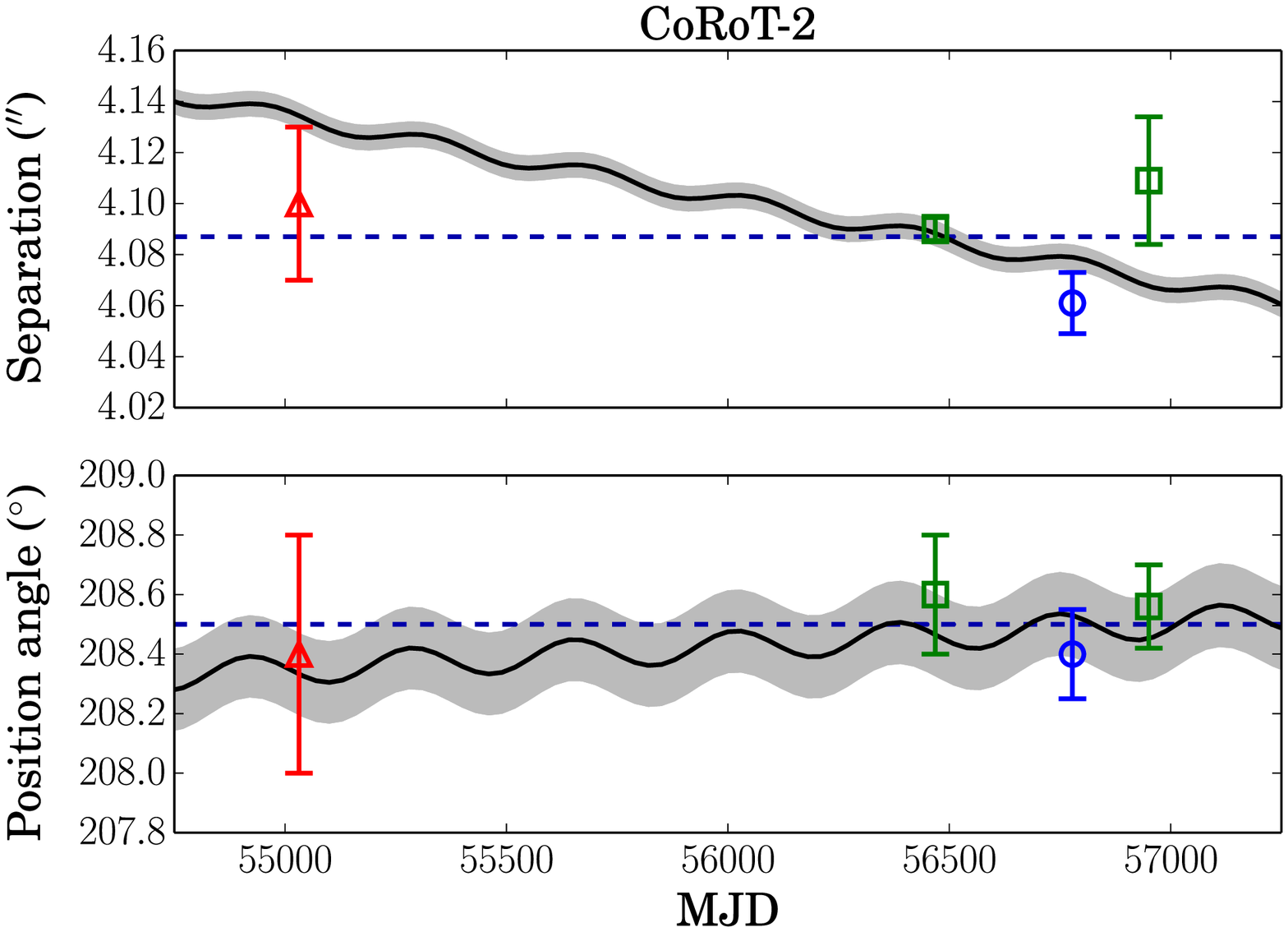} &
		\includegraphics[width=8cm]{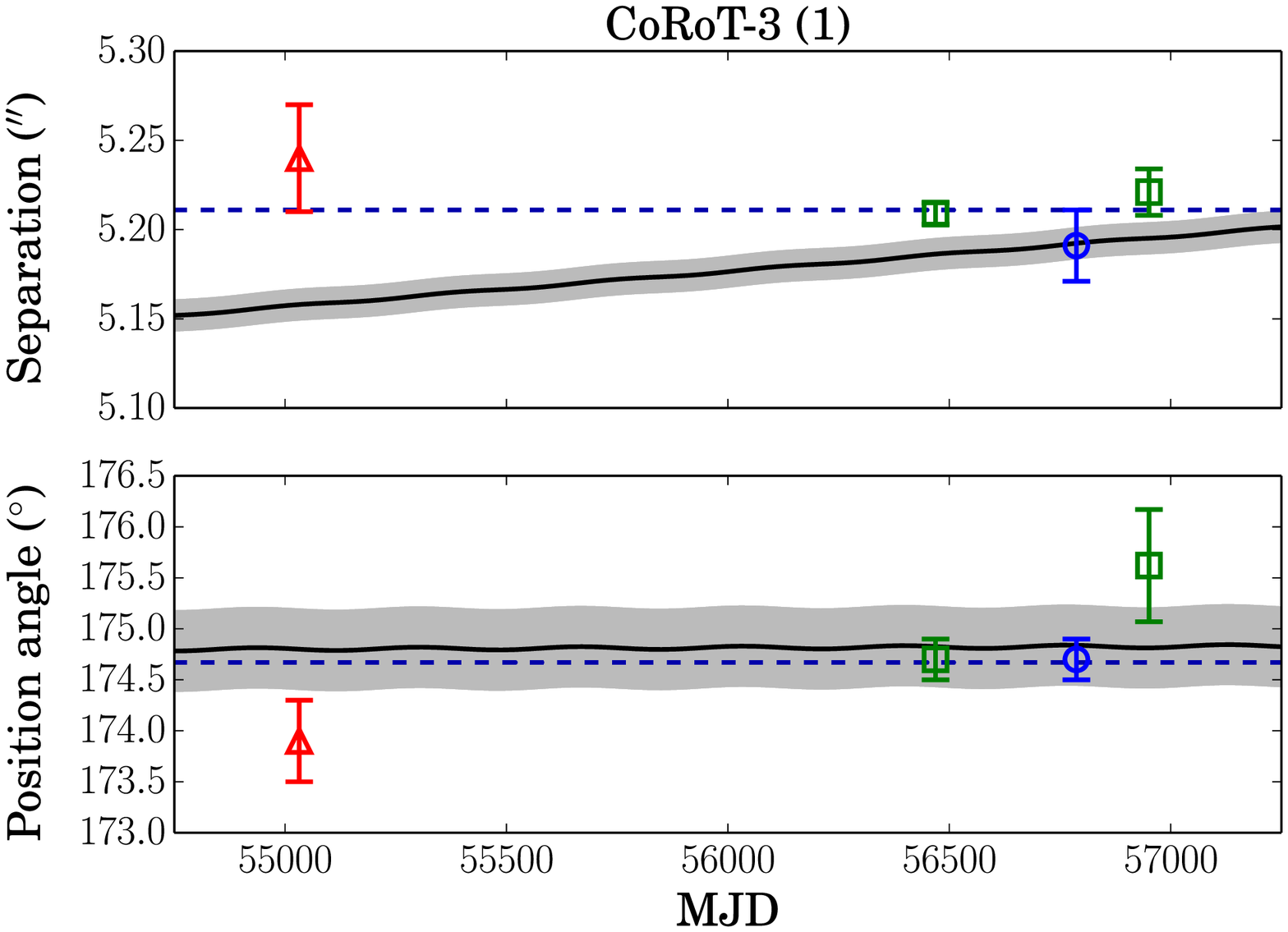} \\
		\includegraphics[width=8cm]{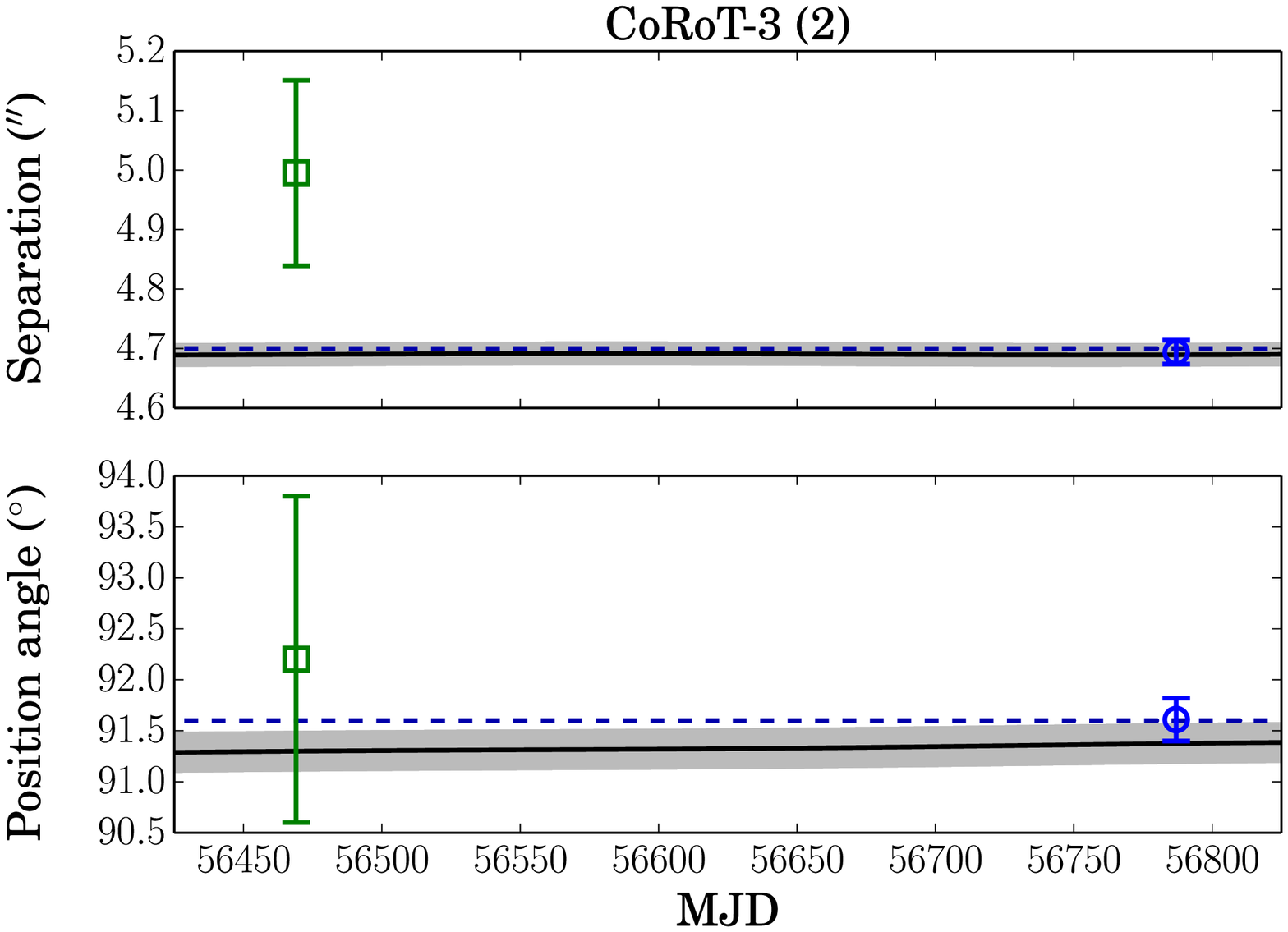} &
		\includegraphics[width=8cm]{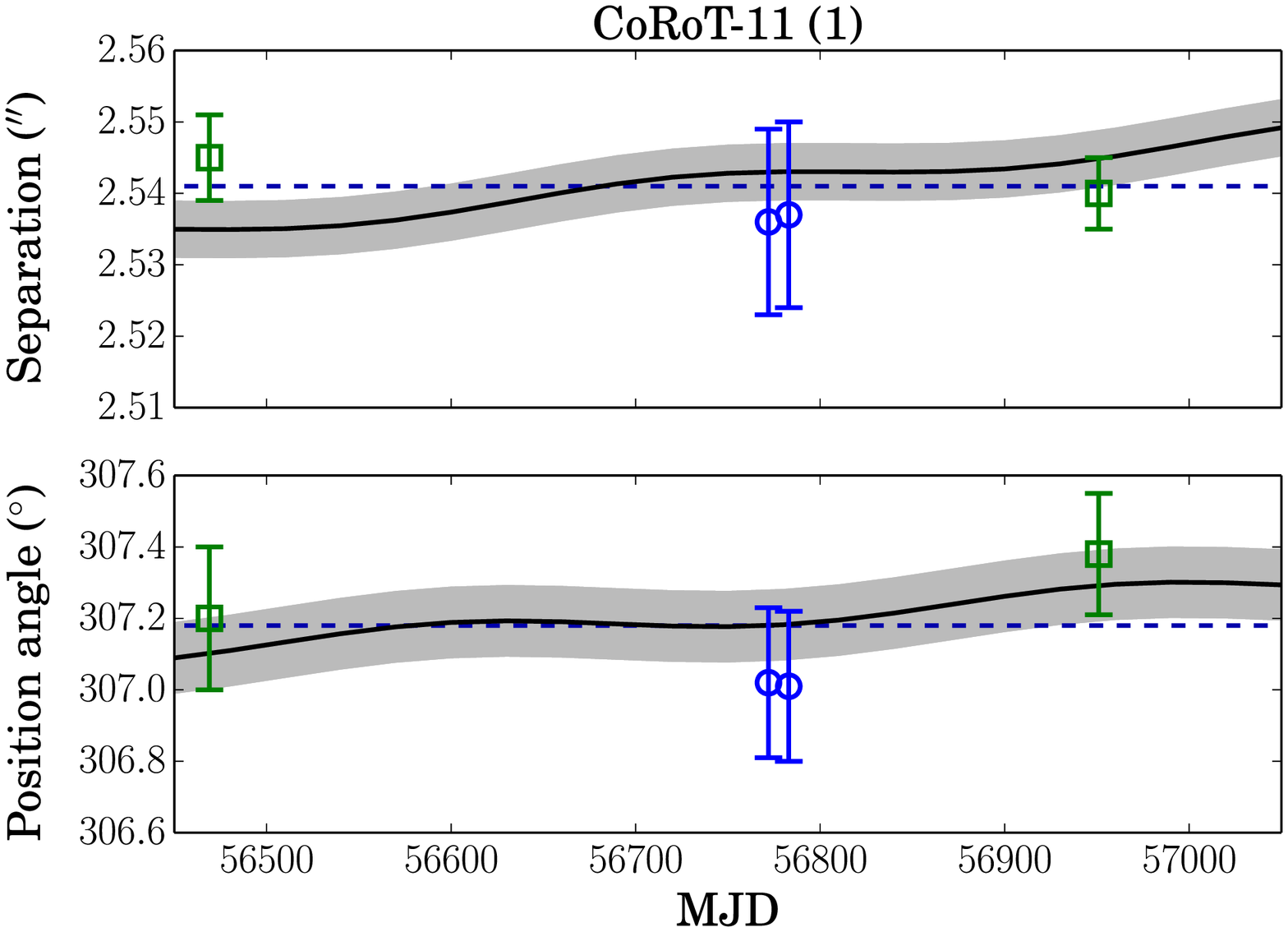}   \\
		\includegraphics[width=8cm]{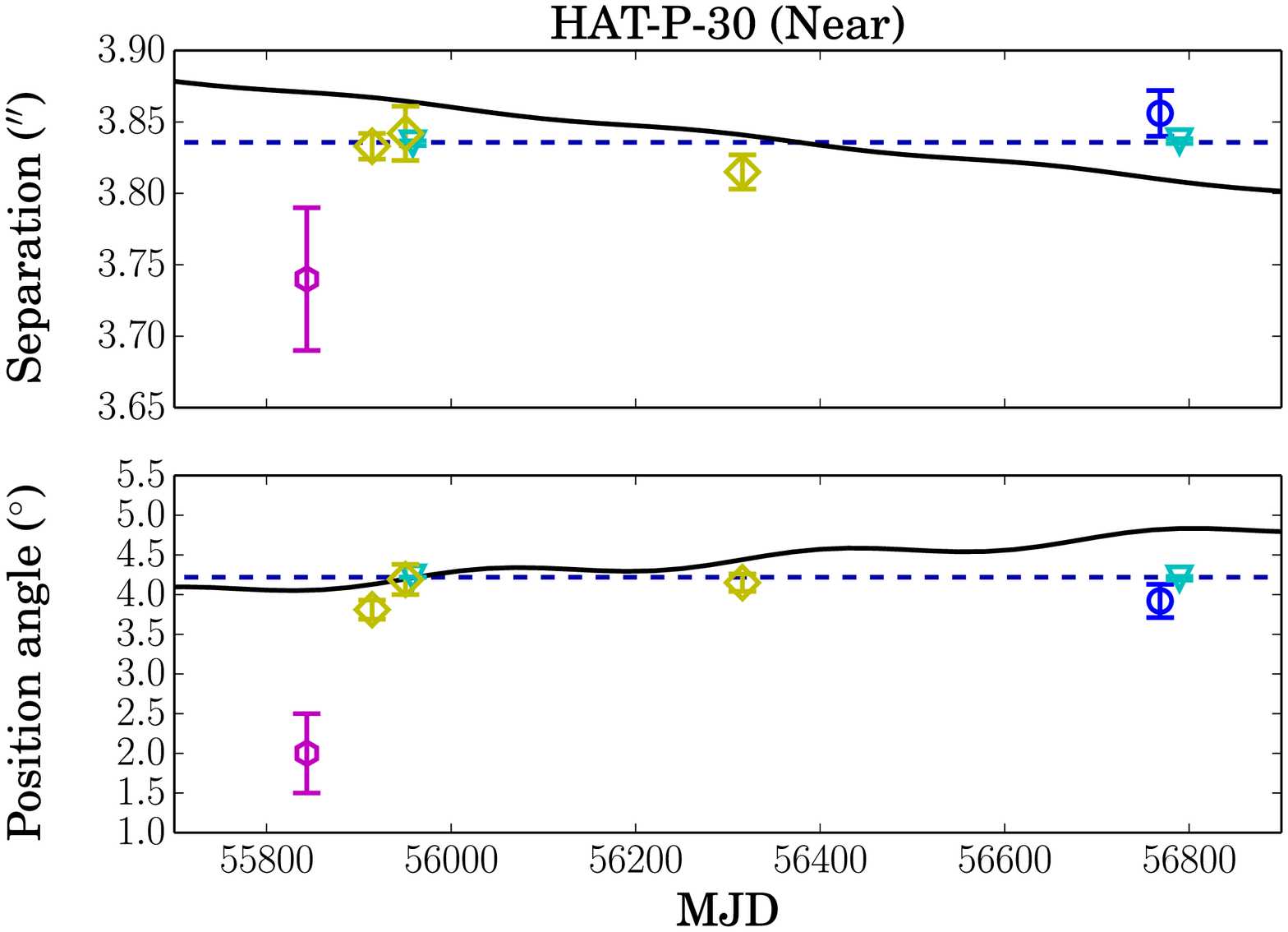} &
		\includegraphics[width=8cm]{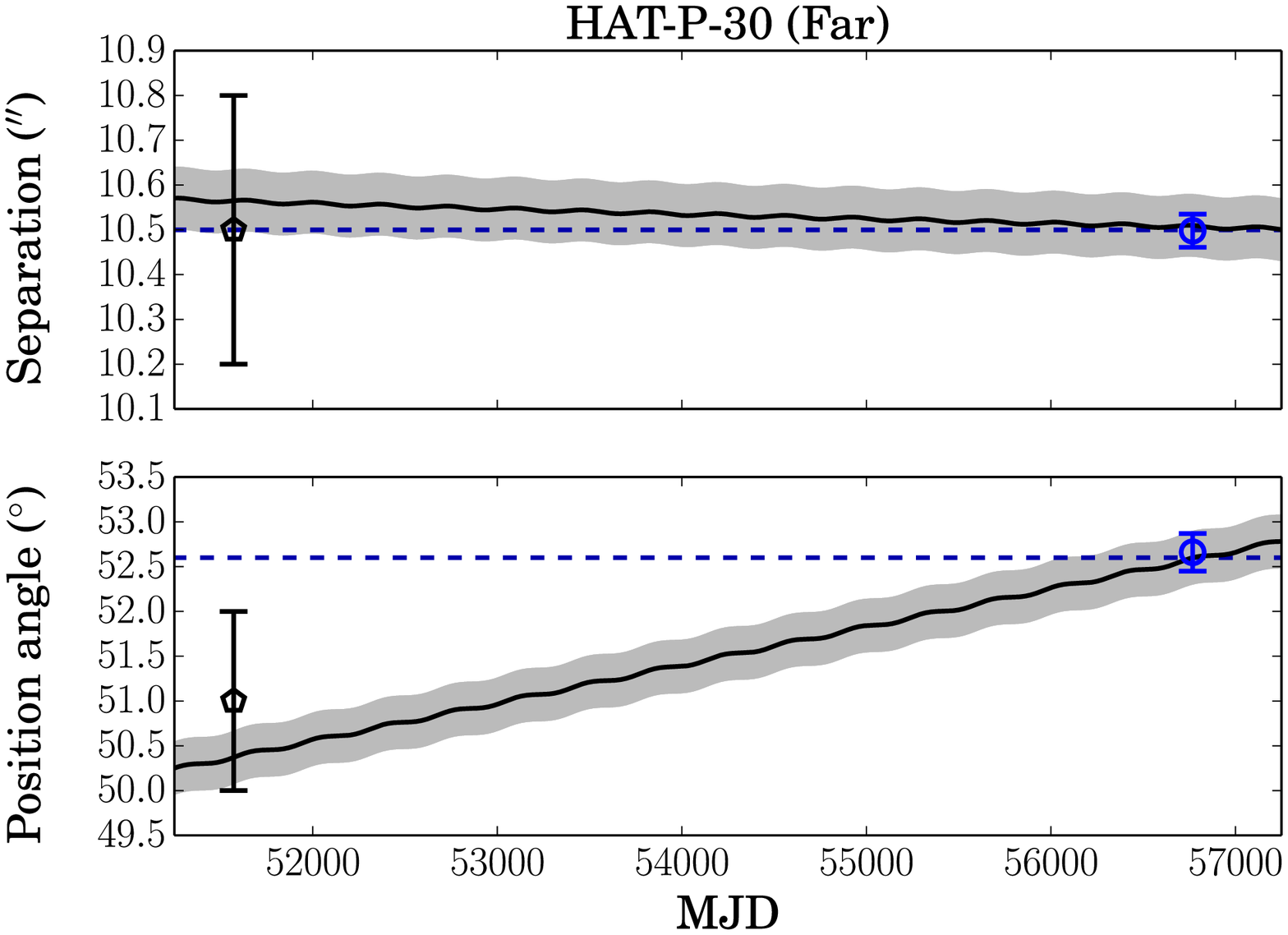} \\
	\end{tabular}
	\caption{\label{fig:cpm1} The relative motions between planet host stars and the candidate companions where observations over several epochs are available. The black lines show the expected motion if the detected star were a distant background star, assuming that such a star would have negligible proper motion compared to the foreground planet host star. The shaded areas indicate the 1$\sigma$ uncertainties on the motions, based on the errors in proper motion. The dashed blue line is the best fit assuming no relative motion between the two stars. The symbols indicate the source of each point: blue circles -- this work; green squares -- \citet{2015A&A...575A..23W, 2015A&A...579A.129W}; red upwards triangles -- \citet{2013MNRAS.433.2097F}; cyan downwards triangles -- \citet{2015ApJ...800..138N}; magenta hexagons -- \citet{2013AJ....146....9A}; yellow diamonds -- \citet{2013A&A...559A..71G}; black pentagons -- 2MASS. } 
\end{figure*}

A nearby star was first reported in \citet{2011AJ....142...86E}, where it was stated that a faint star was observed at a separation of 1.5\as\ during observations with the CORALIE spectrograph, though it is likely that the separation was underestimated (Triaud, 2015, priv. comm.). The system was reobserved in the infrared using adaptive optics in 2011, and a companion was reported at a separation of 3.7\as\ \citep{2013AJ....146....9A}. \citet{2013A&A...559A..71G} reported further measurements of the companion at 3.7\as, finding that it was better explained as a stationary background object than a co-moving companion, whereas \citet{2015ApJ...800..138N} concluded that the two stars do show common proper motion. We report one new observation of the 3.7\as\ companion, and find that it is completely inconsistent with being an unbound background object, as shown in Fig.~\ref{fig:cpm1}, where this companion is labelled `HAT-P-30 (Near)'.


The Washington Double Star Catalog \citep{2001AJ....122.3466M} includes a `C' component for the HAT-P-30 system, observed in 1960 at a separation of 14\as\ and a position angle of 71\degrees, 7\,mag fainter than the A component. We find a star at 10\as\ and a position angle of 53\degrees\ with a similar magnitude difference. The two stars were resolved by the 2MASS survey \citep{2006AJ....131.1163S}, and we derive a separation of 10.5$\pm$0.3\as\ and position angle of 51$\pm$2\degrees\ from the 2MASS observations (MJD at observation 51571.1435), which are shown along with our position measurement in Fig.~\ref{fig:cpm1}, labelled `HAT-P-30 (Far)'. However, our lucky imaging data and the positions from 2MASS do not conclusively prefer either common proper motion or a background star.

We were able to further analyse the more distant candidate companion using its $J-H$ colour index of 0.1$\pm$0.3 derived from the 2MASS photometry, the source being undetected in $K$. This colour characterises the star with a spectral type of A/F if it is on the main sequence \citep{2009BaltA..18...19S}, and hence implies that it is a distant background object given its faintness. We also consider a scenario in which the star is a physically bound white dwarf, using colours and relative magnitudes derived from the evolutionary models presented in \citet{2007A&A...465..249A}, with relative magnitudes in the $I$ filter values being adopted as an approximation to $\textrm{r}_{\textrm{\,TCI}}$. We conclude from the $J-H$ colour and relative magnitudes in $\textrm{r}_{\textrm{\,TCI}}$, $J$, and $H$ that the star is several magnitudes too bright to be a physically bound white dwarf.


\subsection{HAT-P-41}
The HAT-P-41 system was reported to be a potential binary, the planet orbiting an F-type dwarf with a K-dwarf companion 3.56\as\ away \citep{2012AJ....144..139H}. Two sets of observations confirming the presence of the companion have since been published \citep{2015A&A...575A..23W, 2015A&A...579A.129W}. We report a new observation of this companion, denoted `1', and derive a position and magnitude in agreement with previous results. We also report the detection of a new candidate companion star `2' at 1.0\as, 4.4\,mag fainter than HAT-P-41.

As shown in Fig.~\ref{fig:cpm2}, the separation between HAT-P-41 and companion 1 shows an increase over time, which is opposite to the motion that would be shown by a background object, and also much larger than would be expected of orbital motion for such a wide binary. No trend is obvious in position angle, although it should be noted that no measurement was provided in \citet{2012AJ....144..139H}, limiting the observational baseline. We note that the PPMXL and NOMAD catalogues quote significantly different proper motions for HAT-P-41 compared to UCAC4, but these values are also inconsistent with companion 1 either being bound or a stationary background object. One possible explanation for the trend is that the reported companion is a star at a similar or lower distance, exhibiting its own proper motion. Further observations are needed to determine whether or not the trend is real, and if so, its origin.

\begin{table}
	\centering
	\caption{\label{tab:propermotioncomparison} The preferred models of relative motions for candidate companions with previous astrometric measurements. The Common Proper Motion model (CPM) assumes that the two stars have the same proper motion and parallax; the Background model (BG) is based on the candidate companion being a distant background star with negligible proper motion and parallax. Results significant by more than $1.0\sigma$ are highlighted in the table. }
	\begin{tabular}{l l l l }
		\hline \hline
		Target & Comp. & Preferred Model & Significance \\
		\hline
		CoRoT-2 & 1 & CPM & $0.4\sigma$ \\
		CoRoT-3 & 1 & CPM $^1$ & $0.7\sigma$ \\
		CoRoT-3 & 2 & CPM $^2$ & $0.1\sigma$ \\
		CoRoT-11 & 1 & CPM & $0.8\sigma$ \\
		HAT-P-30 & Near & {\bf CPM} & $>5.0\sigma$ \\
		HAT-P-30 & Far & BG & $0.6\sigma$ \\
		HAT-P-41 & 1 & {\bf CPM}& $2.0\sigma$ \\
		WASP-2 & 1 & {\bf CPM}& $>5.0\sigma$ \\
		WASP-8 & 1 & {\bf CPM} & $>5.0\sigma$ \\
		WASP-36 & 1 & CPM & $0.1\sigma$ \\
		WASP-70 & 1 & {\bf CPM} & $>5.0\sigma$ \\
		WASP-77 & 1 & {\bf CPM} & $>5.0\sigma$ \\
		\hline
	\end{tabular}
	\tablefoot{$^1$ NOMAD and PPMXL give significantly different proper motions to UCAC4 for CoRoT-3; the CPM model is preferred at $4.3\sigma$ using NOMAD, and at $2.3\sigma$ PPMXL proper motions. The results for companion 2 are unchanged. $^2$ NOMAD gives a significantly different proper motion to UCAC4 for CoRoT-11; the CPM model is preferred at $3.5\sigma$ using NOMAD.}
\end{table}

\subsection{WASP-2} \label{sec:WASP2}

A companion to WASP-2 was announced in \citet{2007MNRAS.375..951C}, where it was stated to be 2.7\,mag fainter in the $H$ band and located 0.7\as\ to the east. This companion has been further studied in a number of papers using data from the AstraLux Norte and AstraLux Sur lucky imaging cameras. \citet{2009A&A...498..567D} observed the companion in 2007 in the $i'$ and $z'$ filters. \citet{2013MNRAS.428..182B} presented further observations between 2009 and 2011, and showed that the two stars show common proper motion, followed by 2013 data in \citet{2015A&A...575A..23W}. \citet{2013AJ....146....9A} presented $Ks$ band observations from 2011, and $JHK$ observations from 2012/2013 were given in \citet{2015ApJ...800..138N}, along with further analysis supporting the common proper motion hypothesis.

We observed the target three times, but only on 2014-05-10 were the two stars sufficiently resolved to allow the companion to be analysed. The separation and position angle are similar, but not entirely consistent, with previous results. There appears to have been a trend of reducing separation at a rate of $\sim7$ mas/year over a baseline of seven years, illustrated by a red line in Fig.~\ref{fig:cpm2}. This corresponds to a transverse motion of approximately $1.1$ au/year at WASP-2's distance of 140pc \citep{2007MNRAS.375..951C}, which compares to an upper limit of orbital motion of $1.0$ au/year for a $0.85+0.39\Msun$ binary, assuming a circular orbit and that the projected separation is the actual separation. However, given that the two stars appear to be moving directly towards one another, the orbit would have to be nearly edge-on, implying that the projected motion would be much less than $1.0$ au/year when the two stars appear furthest apart. Increasing the orbital radius increases the part of the motion that would be projected, but reduces orbital velocity, and so such an orbit would never result in a projected motion as large as the observed trend.

We obtained a position measurement from the adaptive optics images presented in \citet{2007MNRAS.375..951C} (Skillen, priv. comm.), with a separation of 0.70$\pm$0.03\as\ and a position angle of 101.8$\pm$0.2\degrees on 6/7 September 2007, which does not visibly support the trend. However, even with this additional data, the Bayesian Information Criterion (BIC) of a flat line fitted to the data with no gradient is $638$, compared to a BIC of $26$ for a line including a linear trend, giving a highly significant $\Delta BIC=612$. We note that the fit is mainly constrained by three data points: two astrometric measurements from AO observations by \citet{2015ApJ...800..138N}; and a lucky imaging measurement in 2007 by \citet{2009A&A...498..567D} with a small uncertainty of $0.001\as$. This uncertainty is much smaller than those reported for the observations by \citet{2013MNRAS.428..182B} using the same instrument ($0.024\as$\ and $0.013\as$). However, refitting after excluding the data from \citet{2009A&A...498..567D} still results in a significant result of $\Delta BIC=24$, and the additional exclusion of the precise data from \citet{2015ApJ...800..138N} only reduces this slightly to $\Delta BIC=19$. The fitted trend being in the same direction with a similar magnitude in all cases ($4.9\pm0.3$mas/year for all data, $5.4\pm1.9$mas/year excluding data from \citet{2009A&A...498..567D}, and $7.6\pm3.1$mas/year excluding all high precision data).
	
We therefore conclude that with the currently available data the trend appears to be real. A chance alignment of the motions of two field stars is statistically unlikely due to the high proper motion of WASP-2 ($51.5$ mas/year), which would require the two stars to have very similar proper motions for the trend to be so small. Additionally, all previous analyses of the spectral types of the two stars have found them to be consistent with being at the same distance \citep{2009A&A...498..567D, 2013MNRAS.428..182B, 2015ApJ...800..138N}, although none have considered that the companion may be a distant red giant star. One possible scenario is that the two stars originated from the same star forming region, which could explain the similar proper motions without requiring that they are bound. Further high-precision astrometric data would be useful in confirming the existence of the trend, whilst spectroscopic observations would be able to provide information on the radial motion of the companion, as well as additional information about its physical properties.

\subsection{WASP-8}

\citet{2010A&A...517L...1Q} reported the presence of two stars in the WASP-8 system, with WASP-8B being a faint M-dwarf companion to the planet host star, recorded in both the CCDM and Washington Double Star Catalog. Further observations from 2012 and 2013 were presented in \citet{2015ApJ...800..138N}. We present seven measurements between 2014-07-25 and 2014-09-21, which show little difference to previous results, shown in Fig.~\ref{fig:cpm2}, and find common proper motion is strongly preferred. From two colour photometry, we conclude that the two stars have consistent photometric parallaxes, and derive effective temperatures of 6090$\pm$100K and 3740$\pm$100K for the A and B components respectively. Our value for WASP-8A is higher than the spectroscopically-determined value of 5600K \citep{2010A&A...517L...1Q}, whilst for WASP-8B our values are consistent with previously measured temperatures of 3700K \citep{2010A&A...517L...1Q} and 3380-3670K \citep{2015ApJ...800..138N}. The source of the disagreement in the temperature for WASP-8A is not clear, with our relative magnitudes in $\textrm{v}_{\textrm{\,TCI}}$ and $\textrm{r}_{\textrm{\,TCI}}$ being similar to the differences of $\Delta V = 4.7$ and $\Delta I = 3.5$ measured by \citet{2010A&A...517L...1Q}. An offset in photometrically derived temperatures can be an indicator of an unresolved source, which would have to be bluer than WASP-8A in order to bias our results towards a hotter source. With WASP-8's high proper motion of 111mas/year, it would be expected that a sufficiently bright background source would be visible in previous high resolution observations -- however, we note that no such source was observed by \citet{2015ApJ...800..138N} in 2012 or 2013.

\subsection{WASP-36}

A star separated by 4\as\ from WASP-36 was reported by \citet{2012AJ....143...81S}, with a relative magnitude of 4.8 in the Gunn $r$ filter. The star was observed again in 2015 by \citet{2015A&A...579A.129W}, who found that the candidate companion was dimmer than previously reported, 8.5\,mag fainter than the planet host star in $i$ and 6.5\,mag in $z$. Our position measurements are consistent with those previously reported, and we find that the flux ratio given in \citet{2012AJ....143...81S} agrees well with our value of 4.6\,mag fainter in $\textrm{r}_{\textrm{\,TCI}}$. Our analysis of the motions of the two stars was inconclusive, due to the low proper motion of WASP-36 and the accuracy of the measurements, illustrated in Fig.~\ref{fig:cpm2}.

\subsection{WASP-49}

We present measurements of a candidate companion at 2.2\as\ separation, 5.0\,mag fainter in the red filter. This star was previously observed by chance using the FORS2 instrument in September 2009 by \citet{2015arXiv151206698L},  who found the companion to be $4.3$\,mag fainter in $z'$, with separate spectra of the two stars being used to calculate the relative flux in 27 bins in the range $738$-$1026$nm, showing that the candidate companion is redder. We also confirm the presence of the previously reported star at 9.2\as\ \citep{2012A&A...544A..72L}.

%

%

\subsection{WASP-70}

The WASP-70 system was announced as a G4+K3 binary, with a transiting planet orbiting WASP-70A \citep{2014MNRAS.445.1114A}. The separation was measured as 3.2\as, consistent with archival data and hence indicating no relative motion, and the stars were also found to be co-moving in the radial direction through spectroscopic analysis. Lucky imaging observations from October 2014 have been published by \citet{2015A&A...579A.129W}, and we present an observation from 2014-04-21, shown in Fig.~\ref{fig:cpm2}. We confirm the common proper motion of the two stars, showing it to be significant at the $5 \sigma$ level. 

\subsection{WASP-77}

\begin{figure*} 
	\centering
	\begin{tabular}{cc}
		\includegraphics[width=8cm]{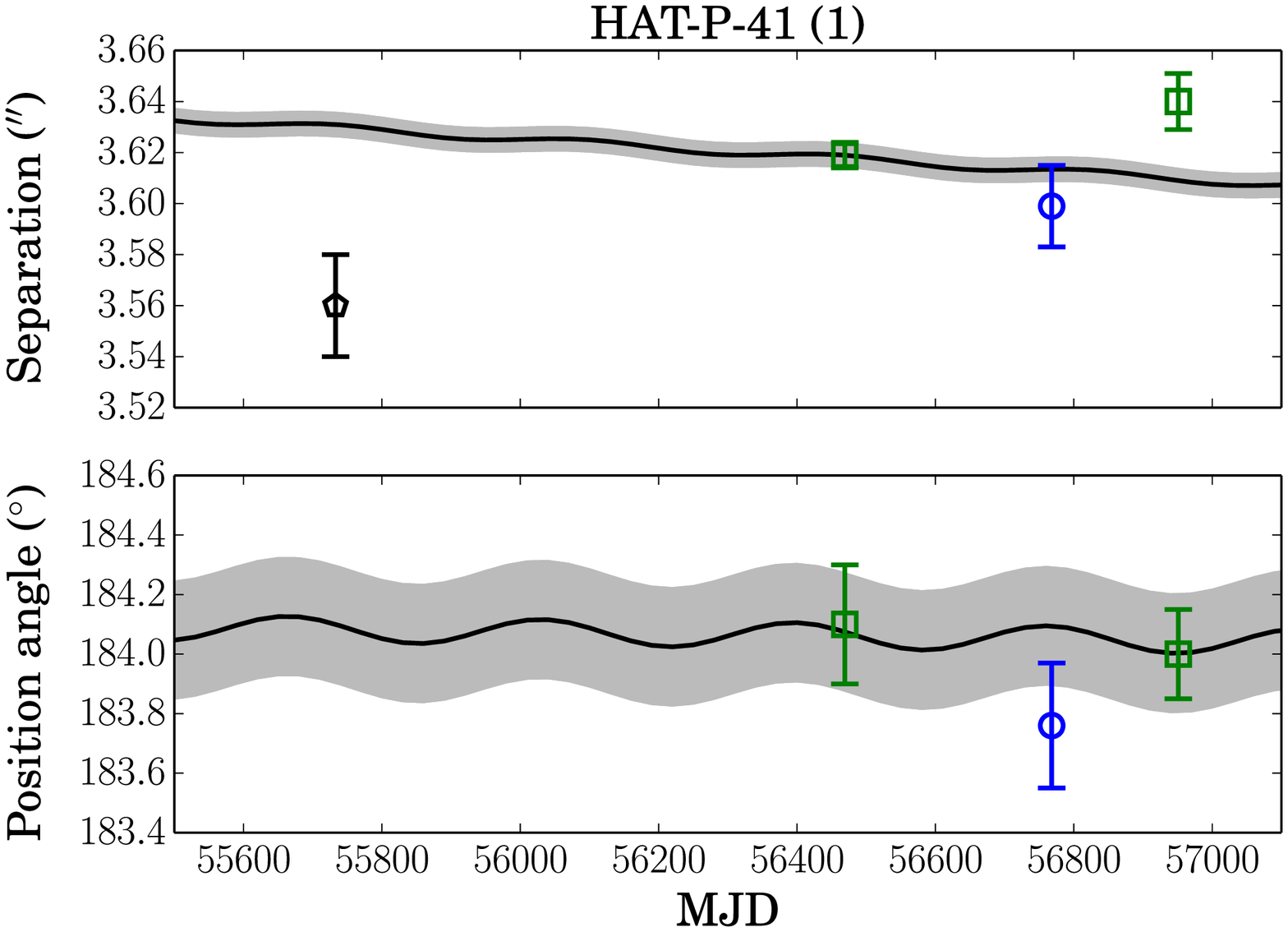} &
		\includegraphics[width=8cm]{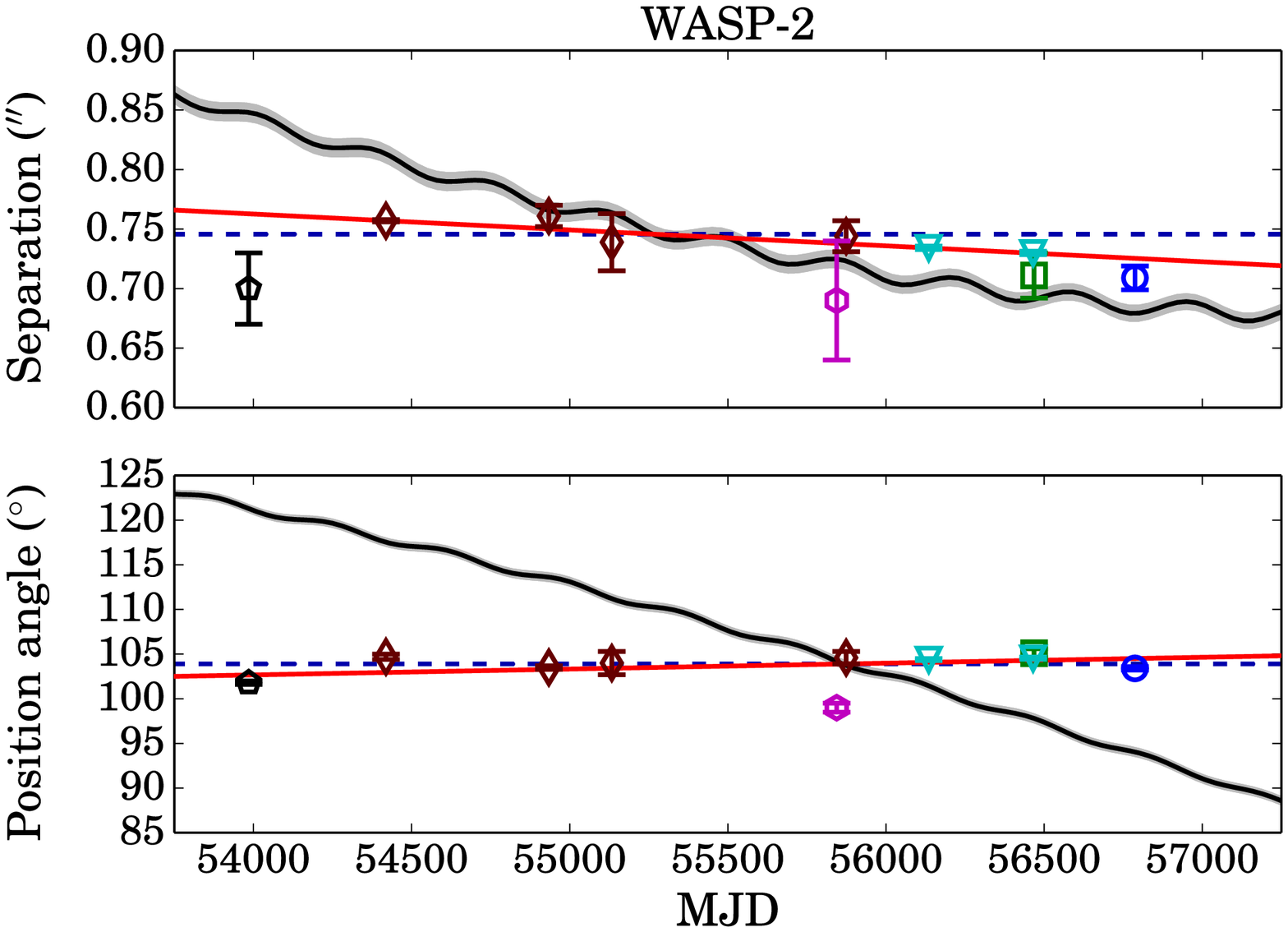}   \\
		\includegraphics[width=8cm]{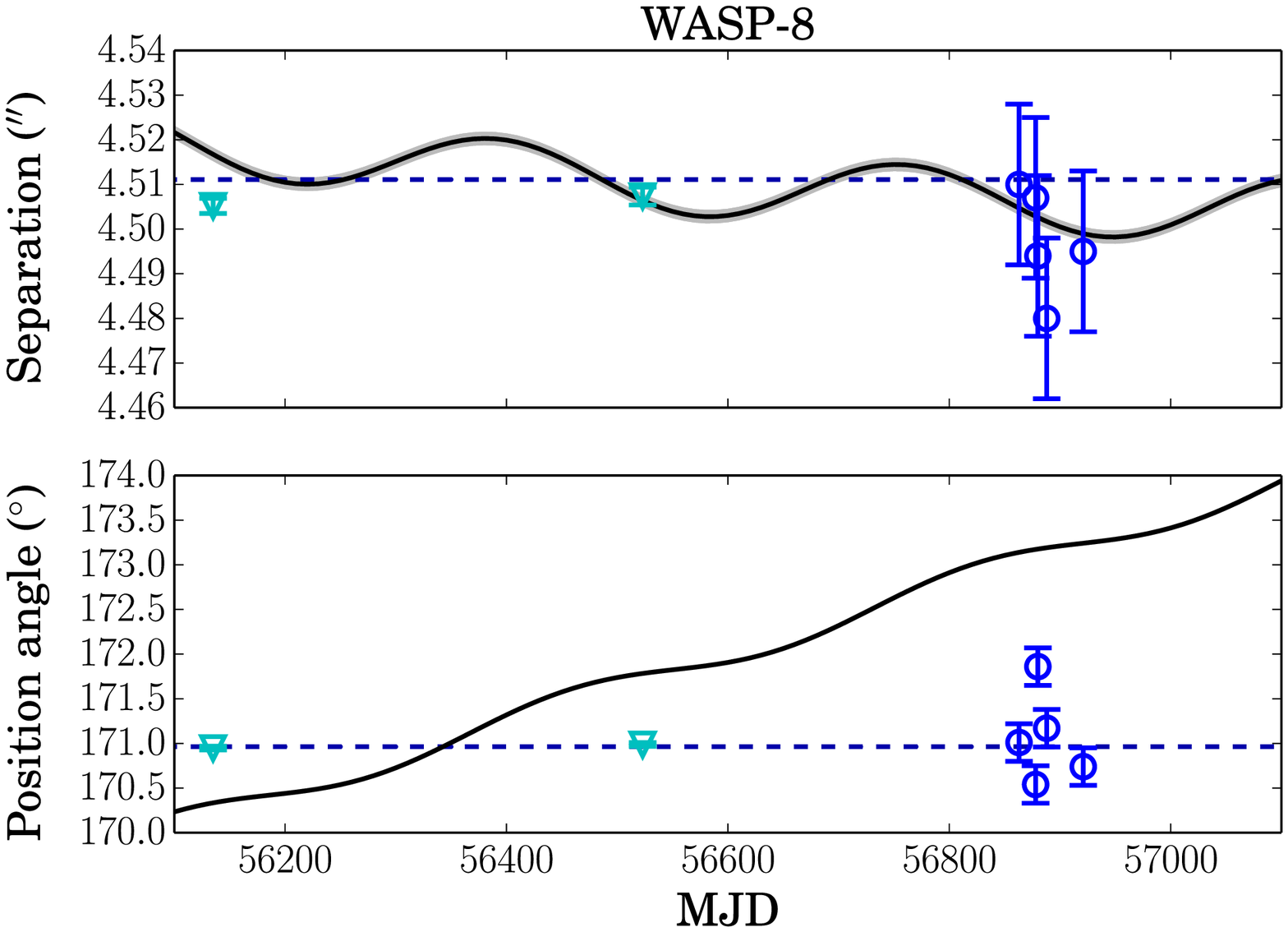} &
		\includegraphics[width=8cm]{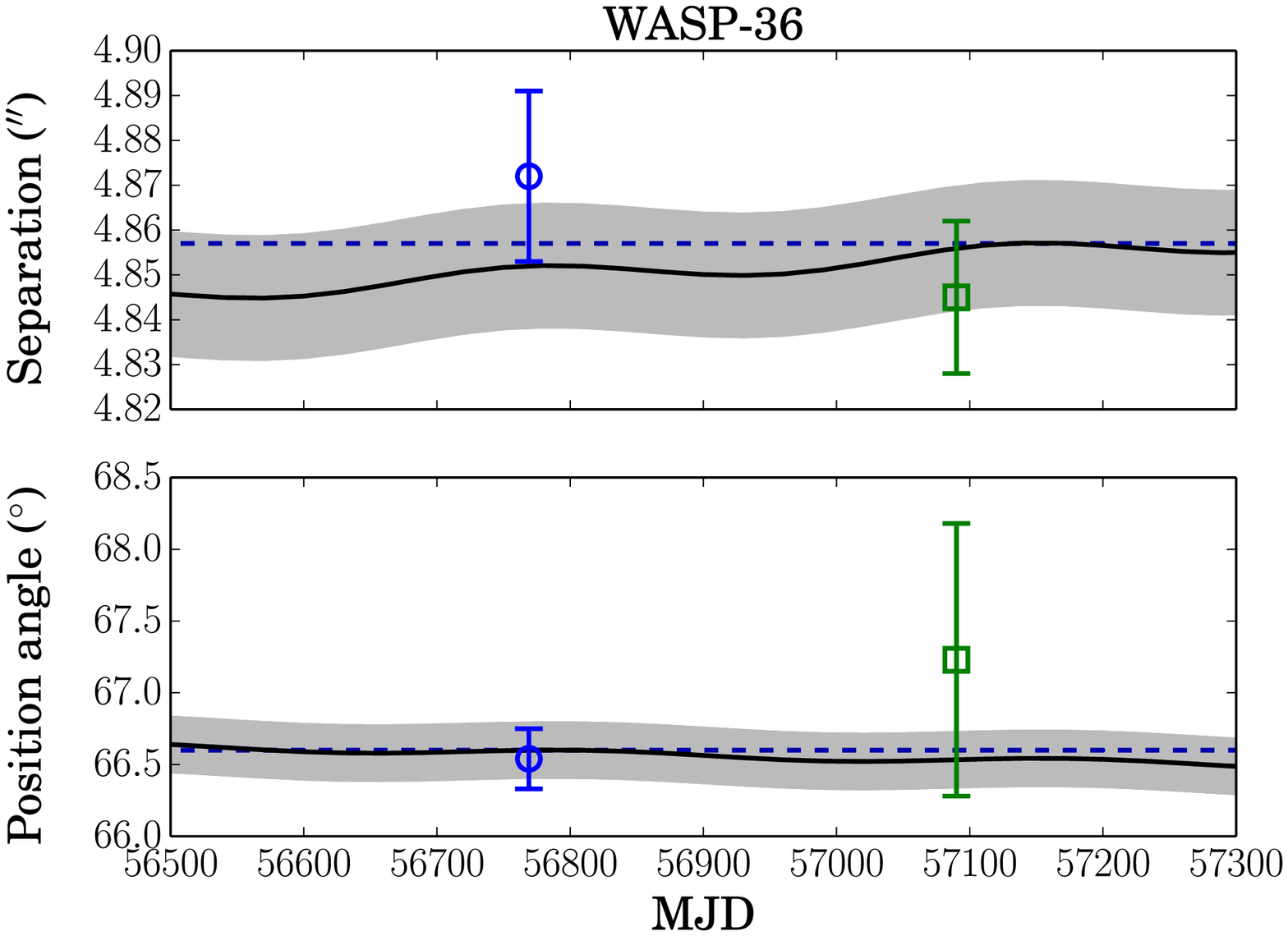} \\
		\includegraphics[width=8cm]{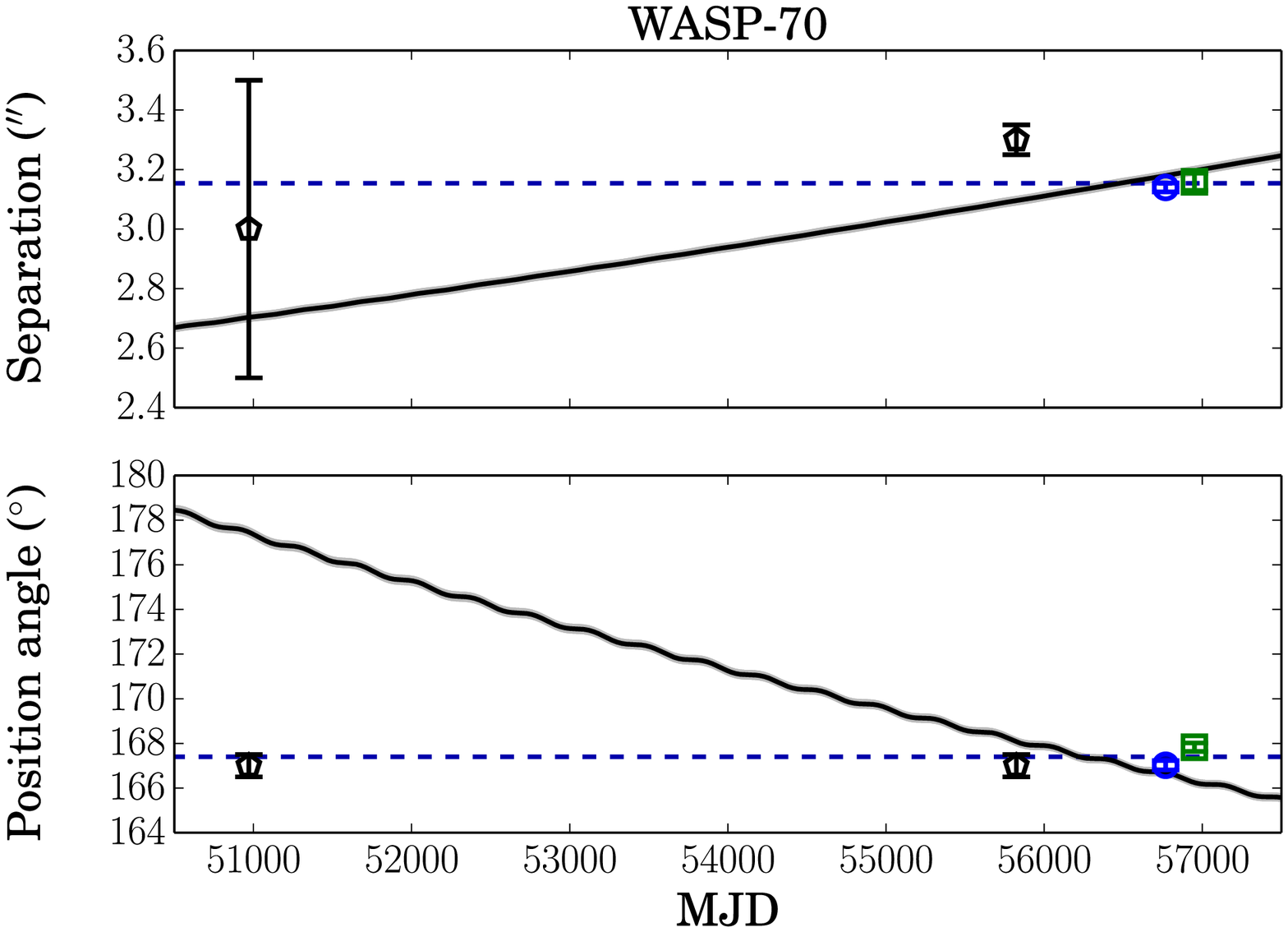} &
		\includegraphics[width=8cm]{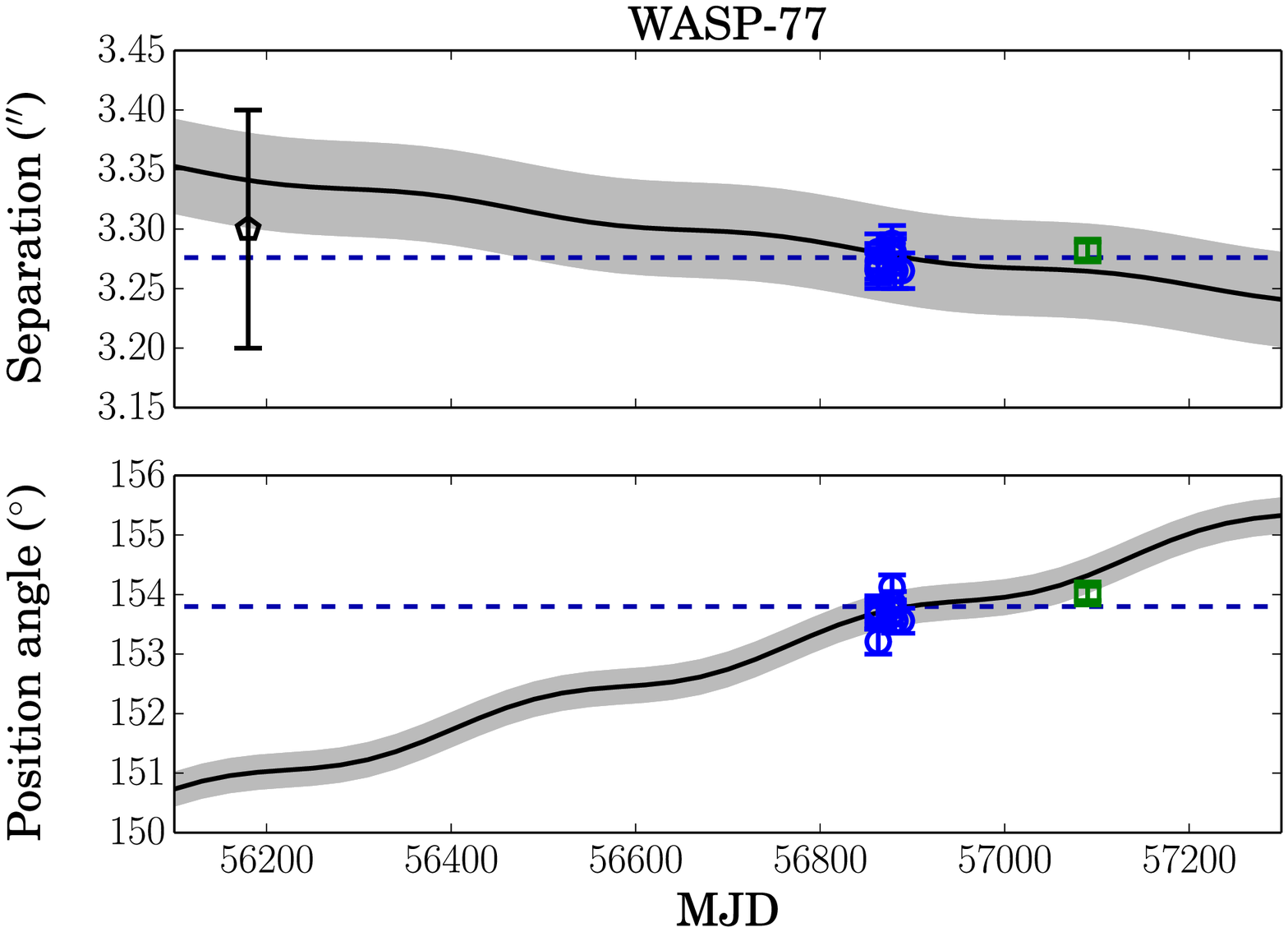}   \\
	\end{tabular}
	\caption{\label{fig:cpm2} The relative motions between planet host stars and the candidate companions where observations over several epochs are available. The black lines show the expected motion if the detected star were a distant background star, assuming that such a star would have negligible proper motion compared to the foreground planet host star. The shaded areas indicate the 1$\sigma$ uncertainties on the motions, based on the errors in proper motion. The dashed blue line is the best fit assuming no relative motion between the two stars. In the case of WASP-2, the red line shows the best fit to the linear trend in separation and position angle, discussed in Section~\ref{sec:WASP2}. The symbols indicate the source of each point: blue circles -- this work; green squares -- \citet{2015A&A...575A..23W, 2015A&A...579A.129W}; cyan downwards triangles -- \citet{2015ApJ...800..138N}; magenta hexagons -- \citet{2013AJ....146....9A}; brown diamonds -- \citet{2009A&A...498..567D, 2013MNRAS.428..182B} black pentagons -- 2MASS, Washington Double Star catalog (WASP-70) or Skillen (priv. comm., WASP-2). } 
\end{figure*}

The WASP-77 system was reported as a G8+K5 binary with a separation of 3.3\as, the planet transiting the primary star \citep{2013PASP..125...48M}. The current positions of the two components are consistent since 1903\footnote{The last two digits of this date were accidentally swapped in \citet{2013PASP..125...48M}}, as recorded in the Washington Double Star Catalog \citep{2001AJ....122.3466M}. \citet{2015A&A...579A.129W} published a lucky imaging observation of the system in October 2014. We present nine observations of the system, with the positions matching well with all previously published values, shown in Fig.~\ref{fig:cpm2}. Common proper motion is clearly preferred at more than $5\sigma$, including the 1903 position measurement in the Washington Double Star Catalog. Using two colour photometry, we derive temperatures of 5830$\pm$100K and 4810$\pm$100K for the A and B components. The temperature of the primary is significantly higher than the value of 5500$\pm$80K given in \citet{2013PASP..125...48M}, whilst our temperature for the secondary is consistent within $3\sigma$ with the previous value of 4700$\pm$100K.



\subsection{WASP-100}

We discover a candidate companion at 4.0\as\ from the planet host star WASP-100, 6\,mag fainter in $\textrm{r}_{\textrm{\,TCI}}$. Using two colour photometry we derive a temperature of 4400K for the star. A physically associated main sequence star with this temperature would be approximately three magnitudes brighter, and hence we conclude that the star is likely a background object.

\subsection{Candidates imaged in two colours}

We list all candidate companions that were imaged and detected in two colours (Red and Visual) in Table~\ref{tab:colourlist}. In each case, the effective temperature of the target star and its companions were estimated from their $\left(\textrm{v}-\textrm{r}\right)_{\textrm{\,TCI}}$ colour using relations interpolated from the data in Table~\ref{tab:colourindices}. The measured colours were corrected for atmospheric extinction and the systematic offset of $-0.46$ mag discussed in Section~\ref{sec:colourcalib}. From the temperatures of each target-companion pair, the magnitude difference that would be expected if the two stars were bound (i.e. at the same distance) was determined from Table~\ref{tab:colourindices}, and the offset between the expected and measured differences was calculated. Based on the scatter in the colour measurements, all temperatures are quoted with a minimum error of $\pm$100K -- this scatter may have been caused by changes to the Visual camera during its commissioning.

The colours of the previously reported bound companion WASP-77 is found to be consistent with this scenario, but that the previously reported companion to WASP-8 is inconsistent -- the temperature that we derive for WASP-8A . Companion 2 to CoRoT-18 is also consistent, and with a calculated distance of 790$\pm$50pc and a projected separation of 11\as, the two stars would be at least 8000 au apart. Several other companions to CoRoT stars are consistent within our uncertainties, but these uncertainties are large and so we do not draw any definite conclusions from this.

We compare the derived temperatures for the planet host stars to the published values contained in the \texttt{TEPCat} database, and find that for the WASP targets and HATS-5 our temperatures are generally within 300K of the expected value. For the CoRoT stars, we find that our values are systematically lower, by up to 1000K -- given the increased distance to these stars and their position in the galactic plane, this systematic error is likely caused by interstellar reddening, which we have not corrected for.

\begin{table*}
	\centering
	\caption{\label{tab:colourlist} The properties of all candidate companions imaged in two colours. The expected relative magnitudes for a bound companion are given in both bands, and the offset between the measured and expected values are shown, divided by the error on the expected values. }
	\begin{tabular}{llccccccc}
		\hline
		\hline
		 &  &  &  &  & \multicolumn{2}{c}{Expected} & \multicolumn{2}{c}{Offset ($\sigma$)} \\
		Target & HJD & Comp. & \Teff\ Targ. (K) & \Teff\ Comp. (K) & $\Delta \textrm{r}_{\textrm{\,TCI}}$ & $\Delta \textrm{v}_{\textrm{\,TCI}}$ & $\Delta \textrm{r}_{\textrm{\,TCI}}$ & $\Delta \textrm{v}_{\textrm{\,TCI}}$ \\
		\hline
		CoRoT-8 & 2456863.6817 & 3 & 4250$\pm$100 & 3410$\pm$130 & 1.9$\pm$0.4 & 2.7$\pm$0.5 & 11.7 & 7.9 \\
		CoRoT-8 & 2456863.6817 & 12 & 4250$\pm$100 & 4200$\pm$190 & 0.0$\pm$0.4 & 0.0$\pm$0.5 & 10.5 & 7.9 \\
		CoRoT-8 & 2456863.6817 & 13 & 4250$\pm$100 & 4500$\pm$400 & -0.5$\pm$0.7 & -0.6$\pm$0.9 & 7.2 & 5.7 \\
		CoRoT-8 & 2456863.6817 & 16 & 4250$\pm$100 & 3500$\pm$1700 & 1.6$\pm$2.8 & 2.2$\pm$3.8 & 1.8 & 1.3 \\
		CoRoT-8 & 2456863.6817 & 17 & 4250$\pm$100 & 3900$\pm$1300 & 0.7$\pm$2.2 & 0.9$\pm$2.9 & 2.4 & 1.8 \\
		CoRoT-8 & 2456863.6817 & 20 & 4250$\pm$100 & 4150$\pm$1500 & 0.2$\pm$2.4 & 0.3$\pm$3.1 & 2.5 & 1.9 \\
		CoRoT-8 & 2456863.6817 & 23 & 4250$\pm$100 & 5300$\pm$1000 & -1.8$\pm$1.3 & -2.3$\pm$1.6 & 5.4 & 4.5 \\
		CoRoT-8 & 2456863.6817 & 24 & 4250$\pm$100 & 4300$\pm$1500 & 0.0$\pm$2.2 & 0.0$\pm$2.8 & 2.7 & 2.1 \\
		CoRoT-8 & 2456863.6817 & 25 & 4250$\pm$100 & 4400$\pm$240 & -0.2$\pm$0.5 & -0.3$\pm$0.6 & 9.7 & 7.5 \\
		CoRoT-8 & 2456863.6817 & 31 & 4250$\pm$100 & 4400$\pm$130 & -0.2$\pm$0.3 & -0.3$\pm$0.3 & 24.0 & 18.4 \\
		CoRoT-8 & 2456863.6817 & 35 & 4250$\pm$100 & 4300$\pm$1000 & 0.0$\pm$1.5 & 0.0$\pm$2.0 & 3.5 & 2.8 \\
		
		CoRoT-18 & 2456924.9054 & 1 & 5270$\pm$100 & 4120$\pm$100 & 2.1$\pm$0.3 & 2.6$\pm$0.3 & 5.6 & 5.6 \\
		CoRoT-18 & 2456924.9054 & 2 & 5270$\pm$100 & 3600$\pm$500 & 3.1$\pm$1.1 & 4.1$\pm$1.5 & 0.7 & 0.5 \\
		CoRoT-18 & 2456924.9054 & 3 & 5270$\pm$100 & 3800$\pm$1300 & 2.7$\pm$2.3 & 3.7$\pm$3.0 & 0.9 & 0.7 \\
		CoRoT-18 & 2456924.9054 & 5 & 5270$\pm$100 & 4100$\pm$1100 & 2.2$\pm$1.8 & 2.7$\pm$2.4 & 1.3 & 1.0 \\
		
		CoRoT-19 & 2456926.8618 & 1 & 5090$\pm$100 & 3400$\pm$1700 & 3.6$\pm$2.9 & 4.9$\pm$4.0 & 1.4 & 1.0 \\
		
		CoRoT-20 & 2456924.8918 & 1 & 5070$\pm$100 & 3000$\pm$1600 & 4.8$\pm$2.8 & 6.7$\pm$3.9 & 0.3 & 0.2 \\
		CoRoT-20 & 2456924.8918 & 3 & 5070$\pm$100 & 3400$\pm$1400 & 3.6$\pm$2.5 & 4.8$\pm$3.4 & 0.7 & 0.5 \\
		CoRoT-20 & 2456924.8918 & 4 & 5070$\pm$100 & 4140$\pm$140 & 1.7$\pm$0.3 & 2.2$\pm$0.4 & 5.2 & 3.8 \\
		CoRoT-20 & 2456924.8918 & 5 & 5070$\pm$100 & 3800$\pm$400 & 2.4$\pm$0.9 & 3.1$\pm$1.2 & 2.2 & 1.6 \\
		
		HATS-5 & 2456879.9143 & 1 & 5520$\pm$100 & 4740$\pm$150 & 1.2$\pm$0.3 & 1.4$\pm$0.4 & 12.7 & 10.3 \\
		HATS-5 & 2456879.9143 & 1 & 5540$\pm$100 & 4730$\pm$140 & 1.3$\pm$0.3 & 1.5$\pm$0.3 & 13.8 & 11.1 \\
		
		WASP-8 & 2456863.9339 & 1 & 6280$\pm$100 & 3840$\pm$100 & 4.0$\pm$0.3 & 4.9$\pm$0.3 & 2.0 & 2.0 \\
		WASP-8 & 2456878.7882 & 1 & 5730$\pm$100 & 3920$\pm$100 & 3.1$\pm$0.3 & 3.9$\pm$0.3 & 1.8 & 1.8 \\
		WASP-8 & 2456880.8132 & 1 & 5930$\pm$100 & 3660$\pm$100 & 4.0$\pm$0.3 & 5.1$\pm$0.4 & 1.9 & 1.9 \\
		WASP-8 & 2456880.8149 & 1 & 6310$\pm$100 & 3780$\pm$100 & 4.1$\pm$0.3 & 5.2$\pm$0.4 & 2.7 & 2.7 \\
		WASP-8 & 2456921.7105 & 1 & 6190$\pm$100 & 3650$\pm$100 & 4.3$\pm$0.3 & 5.5$\pm$0.4 & 3.4 & 3.4 \\
		
		WASP-28 & 2456863.9232 & 1 & 6130$\pm$100 & 3200$\pm$1400 & 5.7$\pm$2.5 & 7.5$\pm$3.0 & 0.5 & 0.4 \\
		WASP-28 & 2456880.7920 & 1 & 6430$\pm$100 & 3280$\pm$140 & 5.6$\pm$0.4 & 7.3$\pm$0.7 & 3.0 & 1.9 \\
		
		WASP-35 & 2456921.8631 & 1 & 5990$\pm$100 & 5600$\pm$300 & 0.5$\pm$0.4 & 0.6$\pm$0.5 & 13.3 & 11.3 \\
		WASP-35 & 2456921.8631 & 2 & 5990$\pm$100 & 3800$\pm$1100 & 3.6$\pm$1.9 & 4.5$\pm$2.6 & 2.0 & 1.5 \\
		
		WASP-77 &2456863.8716 & 1 & 5860$\pm$100 & 4950$\pm$100 & 1.3$\pm$0.2 & 1.5$\pm$0.2 & 1.1 & 1.1 \\
		WASP-77 & 2456863.8730 & 1 & 5830$\pm$100 & 4860$\pm$100 & 1.4$\pm$0.2 & 1.7$\pm$0.3 & 0.5 & 0.5 \\
		WASP-77 & 2456863.8742 & 1 & 5840$\pm$100 & 4880$\pm$100 & 1.4$\pm$0.2 & 1.7$\pm$0.3 & 0.7 & 0.7 \\
		WASP-77 & 2456863.8753 & 1 & 5930$\pm$100 & 4830$\pm$100 & 1.6$\pm$0.2 & 1.9$\pm$0.3 & 0.3 & 0.3 \\
		WASP-77 & 2456878.8952 & 1 & 5770$\pm$100 & 4830$\pm$100 & 1.4$\pm$0.2 & 1.7$\pm$0.3 & 0.2 & 0.2 \\
		WASP-77 & 2456879.8820 & 1 & 5770$\pm$100 & 4670$\pm$100 & 1.7$\pm$0.2 & 2.0$\pm$0.3 & 0.7 & 0.7 \\
		WASP-77 & 2456881.8291 & 1 & 5780$\pm$100 & 4660$\pm$100 & 1.7$\pm$0.2 & 2.0$\pm$0.3 & 0.8 & 0.8 \\
		
		WASP-98 & 2456878.9191 & 1 & 5780$\pm$100 & $>$12000$^1$ &  &  &  & \\
		WASP-98 & 2456879.8985 & 1 & 5740$\pm$100 & 9100$\pm$1900$^2$ &  &  &  &  \\
		WASP-98 & 2456881.8420 & 1 & 5780$\pm$100 & 7200$\pm$1900$^2$ &  &  &  &  \\
		
		WASP-100 & 2456921.8449 & 1 & 6760$\pm$100 & 6800$\pm$400 & -0.1$\pm$0.4 & -0.1$\pm$0.5 & 14.7 & 12.6 \\
		WASP-100 & 2456921.8461 & 1 & 6800$\pm$100 & 6400$\pm$400 & 0.4$\pm$0.4 & 0.4$\pm$0.5 & 13.0 & 11.1 \\      
		\hline      
	\end{tabular}
	\tablefoot{$^1$ The measured $ \left(\textrm{v}-\textrm{r}\right)_{\textrm{\,TCI}}$\ colour of $-0.8$ is outside the range of the colour calibrations presented in Table~\ref{tab:colourindices}, and so an effective temperature could not be calculated. $^2$ The effective temperature of the star is above the range of the Temperature-Radius fit. Therefore, an expected magnitude difference could not be calculated. However, the candidate companion is over 7 magnitudes fainter than the cooler target star, and so is inconsistent with a bound main sequence star. }
\end{table*}

\section{Discussion}                                                                                                           \label{sec:discussion}

\begin{figure} 
	\includegraphics[width=\columnwidth,angle=0]{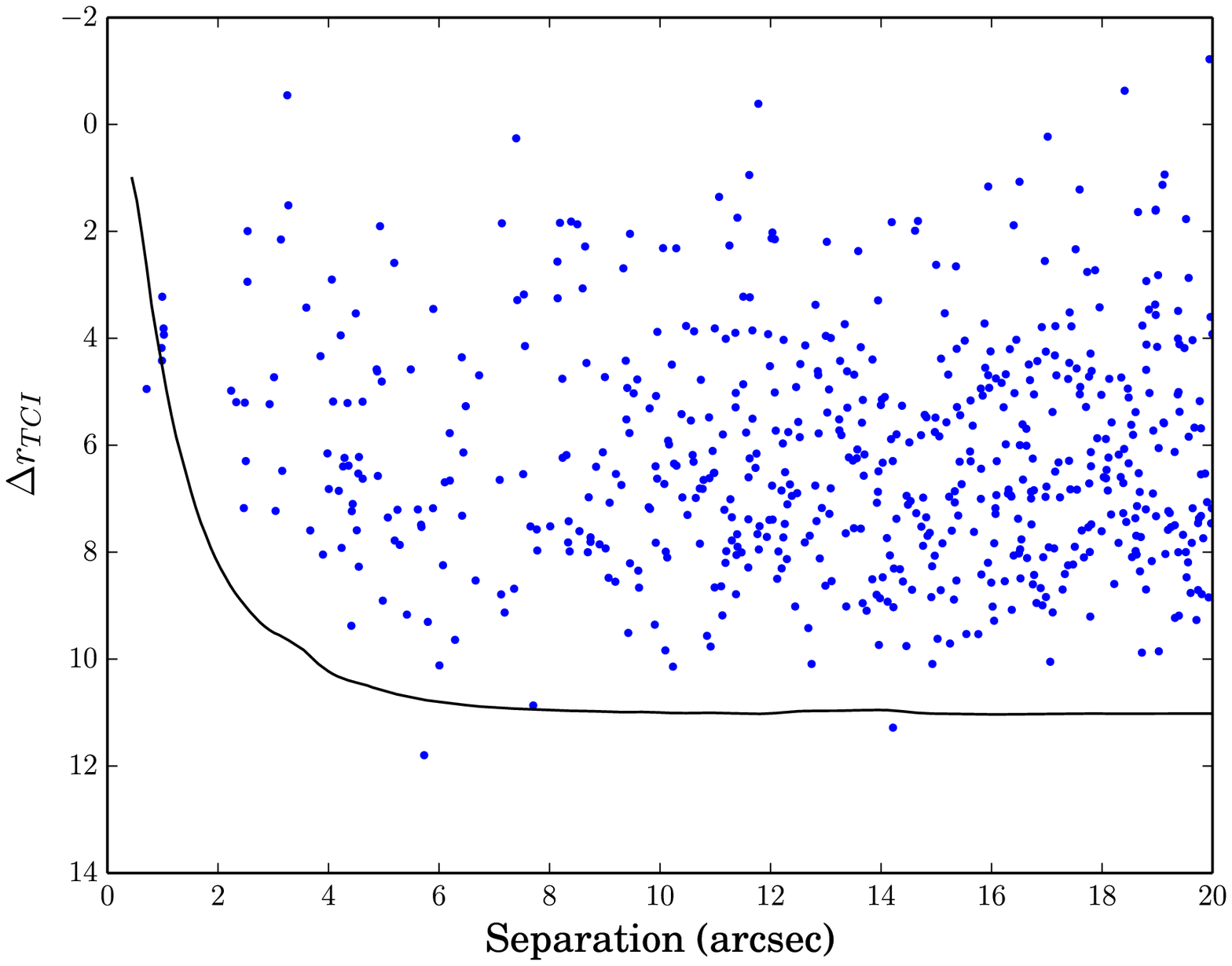}
	\caption{\label{fig:sensitivity} The separation and relative magnitude of all detected candidate companions. Also plotted is the $5\sigma$ detection limit of our observations of WASP-7, typical of our data. Note that the sensitivity of some observations was better at low separation (the closest companion, that to WASP-2, was detected in exceptionally good seeing), and that the lower limit on sensitivity at large separations varied with target brightness and atmospheric conditions. } 
\end{figure}

We report a total of 51 candidate companion stars within 5\as\ of a target star, and 499 are recorded within 20\as, the separation out to which our survey is complete -- these detections are shown in Fig.~\ref{fig:sensitivity}, as well as a typical detection curve from our survey. To analyse the distribution of companions, we divide the host stars into three categories: OGLE host stars, which are situated in very crowded fields towards the galactic centre; CoRoT host stars, which are along the galactic plane, but suffer from less crowding than OGLE targets; and sparse field host stars, including targets such as WASP, HAT, and HAT-South host stars, which are located further away from the galactic plane in less crowded regions of the sky. The number and projected surface density of stars is given in Table~\ref{tab:CompsPerGroup}, divided into the three categories. In all categories the density of stars is higher within 5\as\ than 20\as, as would be expected of a population of bound companions that are more likely to be found at small separations, although the relative increase in density is smaller in the crowded fields, matching the increasing population of background stars. We note that there is a large increase in stellar density for the OGLE stars, especially considering that dim companions are more difficult to detect in crowded fields -- however, the very small sample prevents meaningful analysis of this result.

\begin{table}
	\centering
	\caption{\label{tab:CompsPerGroup} The number of stars, N, found within 5\as\ and 20\as\ for the three groups of targets, and the density of stars per square arcsecond, $\rho_A$. }
	\begin{tabular}{l c c c c c }
		\hline \hline
		Group & Targets & N $\left( 5\as \right)$ & N $\left( 20\as \right)$ & $\rho_A \left( 5\as \right)$ & $\rho_A \left( 20\as \right)$ \\
		\hline
		CoRoT & 13 & 18 & 268 & 0.0176 & 0.0163 \\
		OGLE & 3 & 14 & 168 & 0.0594 & 0.0446 \\
		Other & 85 & 19 & 163 & 0.0028 & 0.0015 \\
		Total & 101 & 51 & 599 & & \\
		\hline
	\end{tabular}
\end{table}

\subsection{Multiplicity Rate}		\label{subsec:multiple}

To estimate the fraction of systems with bound companions, we created a model in which the distribution of detected stars is described using two components. The first is a population of unbound field stars uniformly distributed in space, derived from the \texttt{TRILEGAL} galaxy model, v1.6 \citep{2005A&A...436..895G}. To reduce computation time, the models were generated on a grid with steps of 30\degrees\ in galactic longitude $l$ and 10\degrees\ in galactic latitude $b$, with the grid spacing reduced near the galactic centre ($l>300\degrees$ or $l<60\degrees$, $-10\degrees<b<+10\degrees$) to 10\degrees\ in $l$ and 5\degrees\ in $b$. The default model parameters and magnitude limits ($R=26$) were used, listed in Table~\ref{tab:TRILEGAL}. For each model, 1 square degree of sky was simulated to ensure that a sufficiently large number of stars were generated, with the least dense grid point including over 9,000 stars. The Cousins I band was used as an approximation to $r_{TCI}$. For each target, the density of background stars was based on that of the nearest grid point, the I band magnitude of the target, and the sensitivity curve of our observations.

\begin{table*}
	\centering
	\caption{\label{tab:TRILEGAL} The parameters used to generate the \texttt{TRILEGAL} galactic stellar models. These parameters were the default set for v1.6 of the model at the time of writing. }
	\begin{tabular}{l l }
		\hline \hline
		Parameter & Value \\
		\hline
		Total field area & 1 deg$^2$ \\
		Limiting magnitude & $R=26$ \\
		Distance mod. resolution & 0.1 mag. \\
		IMF & Chabrier Lognormal \\
		Binary fraction & 0.3 \\
		Binary mass ratios & 0.7 to 1.0 \\
		Binary components & Combined (single entry) \\
		Extinction & $A_V(\infty)=0.0378$ \\
		Extinction $1\sigma$ Dispersion & 0 \\
		Sun galactocentric radius & 8,700 pc \\
		Sun height above disc & 24.2 pc \\
		{\bf Thin disc} & \\
		Vertical profile & $\mathrm{sech}^2$ \\
		Scale height & $h=z_0(1+t/t_0)^\alpha$ \\
		$z_0$ & 94.6902 pc \\
		$t_0$ & 5.55079 Gyr \\
		$\alpha$ & $\frac{5}{3}$ \\
		Radial profile & Exponential \\
		Scale length & 2913.16 pc \\
		Cutoffs & 0 pc, 15000 pc \\
		Local calibration $\Sigma_d(\odot)$ & $55.4082\Msun$/pc$^2$ \\
		SFR and AMR & 2-step SFR, Fuhrman AMR, \\
		& $\alpha$ enhanced \\
		SFR and AMR age & $0.73509t$ \\
		{\bf Thick disc} & \\
		Vertical profile & $sech^2$ \\
		Scale height & $h=800$ pc \\
		Radial profile & Exponential \\
		Scale length & 2394.07 \\
		Cutoffs & 0 pc, 15000 pc \\
		Local calibration $\Omega_{td}(\odot)$ & $0.0010\Msun$/pc$^3$ \\
		SFR and AMR & 11-12 Gyr constant, \\
		& Z=0.008, \\
		& $\sigma[M/H]$=0.1 \\
		SFR and AMR age & $t$ \\
		{\bf Halo} & \\
		Shape & Oblate r$1/4$ spheroid \\
		Effective radius $r_h$ & 2698.93 pc \\
		Oblateness $q_h$ & 0.583063 \\
		Local calibration $\Omega_{h} (\odot)$ & $0.000100397\Msun$/pc$^3$  \\
		SFR and AMR & 12-13 Gyr, \\
		& Ryan and Norris $[M/H]$ \\
		SFR and AMR age & $t$ \\
		{\bf Bulge} & \\
		Shape & Triaxial bulge \\
		Scale length $a_m$ & 2500 pc \\
		Truncation scale length $a_0$ & 95 pc \\
		$y/x$ axial ratio $\eta$ & 0.68 \\
		$z/x$ axial ratio $\xi$ & 0.31 \\
		Sun-GC-bar angle $\phi_0$ & 15\degrees \\
		Central calibration $\Omega_b(GC)$ & $406.0\Msun$/pc$^3$ \\
		SFR and AMR & 10 Gyr \\
		& Zoccali et al. 2003 $[M/H]+0.3$ \\
		SFR and AMR age & $t-2.0$Gyr \\
		\hline
	\end{tabular}
\end{table*}

The second model component is made up of bound companions orbiting a fraction $f_{c}$ of the exoplanet host stars, with physical parameters distributed following the results for Solar-type stars in \citet{2010ApJS..190....1R}. The mass ratios $q=\frac{M_2}{M_1}$ of the two components were given a distribution approximating the Solar sample, being randomly placed in one of three regimes: low mass ratio ($0.00<q\leq0.20$, 13.2\% chance), intermediate mass ratio ($0.20<q\leq0.95$, 75\% chance), or high mass ratio ($0.95<q\leq1.00$, 11.8\% chance). This means that components with nearly equal mass were more likely, and systems with extreme mass ratios were less likely. The actual values of $q$ were taken from a uniform distribution of values within the chosen regime. It was assumed that the TEP host star was the more massive star in all cases. The mass of the companion was required to be above the canonical brown dwarf limit of 0.08\Msun\ in all cases, as the relations between temperature, mass, and radius derived in Section~\ref{sec:distance} are not valid in the brown dwarf regime.

The periods follow a log-normal distribution, with a mean $\log\,P$ of 5.03 ($293$ years) and standard deviation of 2.28 (covering $1.5$-$55900$ years), $P$ being expressed in days. Periods were restricted to lie above 10 years ($\log\,P > 3.56$), as a close-in stellar object is likely to have been detected either through radial velocities or by gravitational perturbations of the planet's orbit. Values of eccentricity were uniformly distributed in the range $0.00 \leq e \leq 0.95$, and all other orbital elements, including the phase at time of observation, were distributed uniformly throughout their entire range. The companion ratio, $f_{c}$, varies from 0 to 100\%, i.e. each system may have up to, but not more than, one physically bound companion. Multiple star systems are often found to be of a hierarchical nature, and we assume that a triple stellar system would be composed of the primary orbited at large separation by a close binary, which we count as a single companion. For each randomly generated companion, its projected separation from the planet host star was calculated using the generated orbital elements and the distance to the planet host stars derived in Section~\ref{sec:distance}.

The companion fraction was determined using a likelihood ratio test, with the null model being the special case of $f_c = 0$ (no bound companions), and the alternative model with a non-zero companion fraction. For each target star, we randomly generated 100,000 systems with bound companions and determined their distribution with projected separation, counting any companion that fell below our sensitivity curve as undetected. To determine the expected number of detected companions for each value of $f_c$, the fraction of the simulated systems with detections was multiplied by $f_c$. Therefore, if 50\% of the 100,000 simulated binary systems had a detectable low mass component, and the overall multiplicity fraction $f_c$ was set to 10\%, we concluded that 5\% of stars observed would have a detectable bound companion. The model component consisting of background stars was then added, unmodified by $f_c$.

\begin{figure} 
	\includegraphics[width=\columnwidth,angle=0]{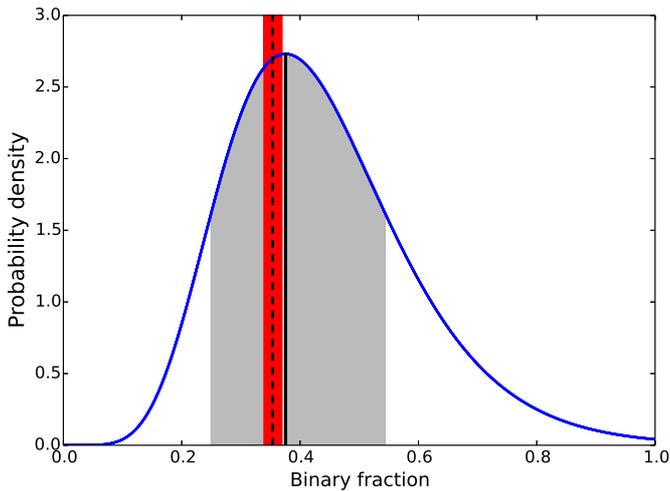}
	\caption{\label{fig:modelfit} The probability density of binary fraction as determined by our statistical model. The black line and grey shaded area indicate the best fit value and 1$\sigma$ uncertainty. The dashed black line and red shaded area show the fraction of Solar-type stars in long period binaries (periods longer than 10 years). } 
\end{figure}

From our model, we find an overall companion fraction of $38^{+17}_{-13}$\% among all target stars. Using the data for Solar type stars in multiple systems from \citet{2010ApJS..190....1R}, the fraction of systems with a component whose period is above 10 years is $35\pm2$\%, consistent with our value. However, a number of long-period systems in the sample of \citet{2010ApJS..190....1R} also include closer companions to the primary, and it has been suggested that planet formation is inhibited in close binaries by theoretical models of planet formation (e.g. \citealt{2012RAA....12.1081Z}). Analysis of the known population of planets has also shown that planets are very rarely found in binaries with separations less than 20 au \citep{2011IAUS..276..409E, 2014ApJ...783....4W, 2015ApJ...806..248W}, with planet frequency being diminished out to binary separations of approximately 100 au \citep{2007A&A...462..345D, 2012A&A...542A..92R}. $28$\% of the systems in the sample of \citet{2010ApJS..190....1R} have a short period component ($P<10$\ years). Removing these systems from the sample and leaving only single stars or multiple systems with long period components only, we find that the expected multiplicity rate among the planet-hosting stars would be $36\pm2$\%. This value is very similar to the fraction among all systems ($35\pm2$\%), indicating that the presence of long period stellar companions is nearly independent of the presence of short period components. As a result, it is not possible to infer any details about any potential lack of short period companions from our survey. The fit to the companion fraction is shown in Fig.~\ref{fig:modelfit}, with the overall multiplicity rate of $35\pm2$\% indicated. 

The consistency of the overall multiplicity rate does not necessarily mean that the distribution of binary properties is the same as the solar sample. \citet{2015ApJ...806..248W} found that whilst the overall stellar multiplicity rate for Kepler gas giant hosts is consistent with solar-type stars, the binary rate is much reduced for the planet host stars below 20 au, enhanced between 20 and 200 au, and finally consistent with the solar sample beyond 200 au. Our survey is mainly sensitive to companions beyond the 200 au limit, and so an enhanced binary rate at smaller separations would not be detected. Further observations of our target systems capable of probing within this limit, such as imaging assisted by adaptive optics or a search for radial velocity trends, are required to determine how the binary fraction changes with distance.

\subsection{Comparison with other surveys}		\label{subsec:comparison}

\citet{2009A&A...498..567D} presented $12\as \times 12\as$ lucky imaging observations of 14 transiting exoplanet hosts finding 3 to have stellar companions, and this survey has since been expanded to cover a total of 31 stars, with 7 potential multiple systems, by \citet{2013MNRAS.428..182B}. A statistical analysis of the whole sample was unable to give stringent limits on the multiplicity fraction, but a lower limit of 38\% was derived \citep{2013MNRAS.428..182B}. Three stars from our survey were observed: WASP-2, with one reported companion, and WASP-7 and WASP-15, for which no companions were discovered. We re-observed the companion to WASP-2 and confirm the lack of any companion to WASP-15, but find a star separated by 4 arcseconds from WASP-7, 9 magnitudes fainter in $r_{TCI}$.

\citet{2013MNRAS.433.2097F} observed 16 systems using lucky imaging out to a radius of 6.5\as, finding 6 companions. Given the low probability of chance alignments with background stars (4\% or less), it was argued that all were likely to be physically bound, giving a multiplicity rate of 38\%. Two targets have been re-observed by our survey: CoRoT-2, with one reported companion, and CoRoT-3, with two companions (although only one was analysed in detail). We re-observe and find consistent results for all companions, as well as finding a number of additional faint companions for CoRoT-3.

12 Kepler targets and 15 TEP hosts were observed by \citet{2013AJ....146....9A} using adaptive optics. Of the non-Kepler systems, companions were reported within 4\as\ for five stars. Two of the systems with companions, WASP-2 and HAT-P-30, have been re-observed by our survey and several others, although we note that the results in \citet{2013AJ....146....9A} appear to suffer from systematic errors in both separation ($\sim$5\% too low) and position angle ($\sim$2-5 degrees). We also confirm the lack of any companion within 4\as\ of CoRoT-1.

A survey of 50 systems using adaptive optics was presented by \citet{2015ApJ...800..138N} with a $10\as\times10\as$ field of view. Statistical analysis resulted in an estimated companion rate of $49\pm9$\%, in agreement with our finding of $38^{+17}_{-13}$\%. Twelve targets were re-observed in this survey, of three were reported to have a companion, and in all cases we confirm the presence (or lack of) a close companion.

\citet{2015A&A...575A..23W} and \citet{2015A&A...579A.129W} presented a large lucky imaging survey of TEP host stars, with a field of view of $12\as\times12\as$. 37 targets were also reobserved with our survey, of which 12 were reported to have companions. For ten of these systems, we confirm the existence of the companions. For the remaining two systems, HAT-P-27 and WASP-103, the companions are below our sensitivity curve, and we are unable to detect them. For the CoRoT-3, CoRoT-7, CoRoT-11 and WASP-90 systems, we find a total of 7 faint companions within 5\as\ that were not detected by \citet{2015A&A...575A..23W} and \citet{2015A&A...579A.129W}. Whilst the exposure times are similar to those in our survey, ranging from 150s to 900s, we note that only the best 10\% of lucky imaging exposures were used, rather than the maximum fraction of 90\% used in this paper. As a result, our sensitivity at large separations is 2-3 magnitudes better for most targets.



\section{Conclusions}                                                                                                                \label{sec:conc}

We present lucky imaging observations of 101 TEP host systems in the southern hemisphere, with measurements of 499 candidate companion stars within 20\as\ of a target: 168 stars were found near the 3 OGLE targets located along the galactic plane, 268 near 13 CoRoT targets, and 163 near the remaining 85 targets. The majority of these stars are likely to be unbound background objects. Within 5\as, 51 stars were found -- due to their smaller separation, these stars are more likely to be physically related to the target stars, and also provide the greatest amount of photometric and spectroscopic contamination.

For 12 candidate companions with historical observations we present proper motion analyses, and we also provide two band photometry for a subset of the targets. We find that a star located 10 arcseconds from HAT-P-30, recorded as a bound companion in the Washington Double Star Catalog, is likely being a background star. For the close companion to HAT-P-30, and the companions to WASP-8, WASP-70 and WASP-77, we confirm common proper motion. For WASP-2B, we provide a revised separation and position angle for the 2007 observation that better matches later measurements. We find that the two stars in the WASP-2 system appear to be moving together on the sky, but find a statistically significant trend of reducing separation of approximately $0.8$ mas/year corresponding to a projected motion of $1$ au/year, far above the expected orbital motion. Additionally, no such trend is apparent in position angle, and we are unable to explain the origin of the trend -- possible scenarios include a chance alignment of two stars at a similar distance, similar proper motions of two stars originating from the same starforming region, or that the companion is actually a distant red giant star with a high velocity. For the previously reported companion to HAT-P-41, we find evidence that the two stars are moving relative to one another, suggesting that the two stars are not physically related. The remainder of our proper motion analyses are not conclusive, due to low numbers of observations, low proper motions of the target stars, or inconsistent measurements from different groups.

We test our distribution of candidate companions against a model which includes both physically bound and background stars. By comparing the number and separation of companions against a varying companion fraction, we conclude that hot Jupiter host stars have a multiplicity rate of 38$^{+17}_{-13}$\%, consistent with the fraction of Solar-type stars that have a companion star with an orbital period over 10 years (35$\pm$2\%). We also investigate the effect of close binaries, which are known to suppress planet formation -- many stellar multiple systems are hierarchical, containing both short period and long period components. We find that the expected multiplicity rate if planet formation occurs around only single stars or those in long period binaries is 36$\pm$2\%, almost indistinguishable from the fraction expected if systems with short period components are included. We therefore conclude that our sample does not show an overabundance of long period companions in the range of 1-40\as, suggesting that planet formation does not preferentially occur in wide stellar binaries. The populations of companions at short and long periods do not appear to be linked, and so we are unable to make any inferences about the presence -- or lack of -- short period stellar companions. Probing this regime requires powerful imaging techniques such as adaptive optics imaging, with the latest high contrast instruments such as SPHERE \citep{2008SPIE.7015E..18M} or GPI \citep{2008SPIE.7014E..18B} being best suited to finding dim stellar companions at distances less than a few tens of au, where the dynamical influence of such companions would be strongest. Alternative techniques sensitive to close-in companions should also be considered, with examples including radial velocity surveys (e.g. \citealt{2014ApJ...785..126K}) or spectroscopic searches for unresolved stars (e.g. \citealt{2015ApJ...814..148P}).

The results of calibrations for the detector scale and the colour response of the TCI are presented, along with a set of theoretical stellar colour indices. Since late 2014, the TCI has been capable of simultaneous two colour photometry, and multicolour observations of candidate companions allow us to determine whether or not their distance is consistent with that of the target star, providing an efficient way to detect background objects. We analyse the stars for which two colour photometry was available, and find that the bound companions to WASP-8 and WASP-77 are consistent with being at the same distance. We compare the derived effective temperatures for the planet host stars with two colour photometry, and find them generally reliable, but that systematic offsets can be caused by interstellar extinction, for which a correction is not yet available.

In the future we will obtain new measurements of the separation and position angle of candidate companions, allowing the detection of common proper motion. Two colour photometry will be performed for all future observations, allowing the relative distances to the target and companion stars to be calculated, providing an additional indicator of boundedness. We will also be extending the HITEP campaign to include analyses of several recently published transiting exoplanet systems.


\begin{acknowledgements}
	D.F.E. is funded by the UK’s Science and Technology Facilities Council. J. Southworth acknowledges support from the Leverhulme Trust in the form of a Philip Leverhulme prize. D.B. acknowledges support from NPRP grant \# X-019-1-006 from the Qatar National Research Fund (a member of Qatar Foundation). G.D. acknowledges Regione Campania for support from POR-FSE Campania 2014-2020. T.C.H. acknowledges KASI research grant 2014-1-400-06. N.P. acknowledges funding by the Gemini-Conicyt Fund, allocated to the project No. 32120036. J. Surdej and O.W. acknowledge support from the Communant\'e fran\c caise de Belgique -- Actions de recherche concert\'ees -- Acad\'emie Wallonie-Europe. The operation of the Danish 1.54 m telescope is financed by a grant to U.G.J. from the Danish Natural Science Research Council (FNU). We acknowledge the use of the NASA Astrophysics Data System; the SIMBAD database and the VizieR catalogue access tool operated at CDS, Strasbourg, France; and the arXiv scientific paper preprint service operated by Cornell University. This paper makes use of data obtained from the Isaac Newton Group Archive which is maintained as part of the CASU Astronomical Data Centre at the Institute of Astronomy, Cambridge. This publication makes use of data products from the Two Micron All Sky Survey, which is a joint project of the University of Massachusetts and the Infrared Processing and Analysis Center/California Institute of Technology, funded by the National Aeronautics and Space Administration and the National Science Foundation.
\end{acknowledgements}


\bibliographystyle{aa}
\bibliography{dfe}


\end{document}